# Long-Range Exciton Diffusion in Two-Dimensional Assemblies of Cesium Lead Bromide Perovskite Nanocrystals

Erika Penzo,[1]* Anna Loiudice,[2] Edward S. Barnard,[1] Nicholas J. Borys,[1†] Matthew J. Jurow,[1,3]

Monica Lorenzon,[1] Igor Rajzbaum,[1] Edward K. Wong,[1] Yi Liu,[1,3] Adam M. Schwartzberg,[1]

Stefano Cabrini,[1] Stephen Whitelam,[1] Raffaella Buonsanti,[2] Alexander Weber-Bargioni[1]*

[1] The Molecular Foundry, Lawrence Berkeley National Laboratory, Berkeley, CA 94720, USA

[2] Institute of Chemical Sciences and Engineering of the École Polytechnique Fédérale de Lausanne, CH 1015, Switzerland

[3] Materials Sciences Division, Lawrence Berkeley National Laboratory, Berkeley, CA 94720, USA

**Corresponding Authors**

* Erika Penzo: erikapenzo@gmail.com

*Alexander Weber-Bargioni: awb@lbl.gov

ABSTRACT

Förster Resonant Energy Transfer (FRET)-mediated exciton diffusion through artificial nanoscale building block assemblies could be used as an optoelectronic design element to transport energy. However, so far nanocrystal (NC) systems supported only diffusion lengths of 30 nm, which are too small to be useful in devices. Here, we demonstrate a FRET-mediated exciton diffusion length of 200 nm with 0.5 $cm^2$/s diffusivity through an ordered, two-dimensional assembly of cesium lead bromide perovskite nanocrystals ($CsPbBr_3$ PNCs). Exciton diffusion was directly measured *via* steady-state and time-resolved photoluminescence (PL) microscopy, with physical modeling providing deeper insight into the transport process. This exceptionally efficient exciton transport is facilitated by PNCs' high PL quantum yield, large absorption cross-section, and high polarizability, together with minimal energetic and geometric disorder of the assembly. This FRET-mediated exciton diffusion length matches perovskites' optical absorption depth, thus enabling the design of device architectures with improved performances, and providing insight into the high conversion efficiencies of PNC-based optoelectronic devices.

Energy transport at the nanoscale plays a critical role in a plethora of natural systems: for instance, the photosynthetic process relies on Förster Resonant Energy Transfer (FRET) to transport energy along a few μm of diffusion length,[1] as well as being the primary mechanism of energy transport between proteins or different moieties of the same protein.[2] A detailed understanding of the FRET mechanism is pivotal to the realization of artificial systems with efficient, long-range energy propagation.

Quantum dots (QDs) solids have recently gained a lot of attention, thanks to the advantages of engineering 3D arrays with nanoscale building blocks, the QDs, which are well known for their exceptional optical properties, as well as for the possibility to easily tune such properties by changing their size, composition and surface chemistry.[3-5] In addition, their self-assembly in close-packed systems facilitates the communication between neighbouring QDs, by enabling FRET of excitons that are able to hop onto adjacent, non-excited QDs, which results in the transport of the excitonic energy for multiple steps before the exciton recombines. Understanding FRET-mediated exciton diffusion is critical for enhancing the performances of optoelectronic devices, such as flexible organic light-emitting diodes (OLEDs), solar cells, and light modulators,[6] by engineering the diffusion appropriately for the desired application. For instance, solar cells largely benefit from the migration of excitons towards the charge-separation interfaces.[5, 7] Conversely, in light-emitting devices a large diffusion of exciton is detrimental to the efficiency, since it prevents the exciton from radiatively emitting from the QD where it formed, thus risking his non-radiative trapping onto adjacent layers.[5,7] Artificial nanoscale building block assemblies are characterized by relatively short FRET-mediated exciton diffusion lengths, typically on the order of 10 nm in organic semiconductors,[8] 30 nm in inorganic nanocrystal (NC) solids,[9] and up to 50-70 nm in perovskite nanocrystals (PNCs) assemblies,[10] which is a major limitation to the exploration of exciton-based optoelectronic phenomena and to the development of optoelectronic devices. Here we show that, within close-packed, two-dimensional assemblies of isoenergetic PNCs, FRET-mediated exciton diffusion lengths reach 200 nm, close to the light absorption depth of 200-400 nm for this class of materials,[11] and the so far longest reported FRET-mediated exciton diffusion length in a NC system.

The Förster equation describes the structural and optoelectronic requirements between a donor-acceptor system to maximize the FRET rate:[12]

$$^1\!/_{\tau_{DA}} = \frac{2\pi}{\hbar}\frac{\mu_D^2\mu_A^2\kappa^2}{\mathrm{r}_{DA}^6 n^4}$$

$^1\!/_{\tau_{DA}}$ is the hopping rate between Donor (D) particle and Acceptor (A) particle with $\tau_{DA}$ the average time for an exciton to hop from donor to acceptor. Maximizing this hopping rate requires (1) minimizing inter-particle distances *r* and the refractive index *n,* while (2) maximizing the spectral overlap between donors and acceptors $\mu_D^2\mu_A^2$, and (3) aligning the dipole moments to maximize the orientation factor $\kappa^2$. Additional parameters critical to experimentally study the transport are maximizing the polarizability of individual emitters to enhance (4) photon absorption and hence exciton creation, (5) the fluorescent quantum yield for an optimized signal-to-noise ratio, and (6) a flat energy landscape between the particles (small inhomogeneous line broadening). Importantly, these 6 parameters affect the exciton diffusion multiplicatively, meaning that if one parameter is poorly optimized, exciton diffusion can be completely suppressed. Hence, most studies of exciton transport through nanoscale building block assemblies find exciton diffusion length of 6-30 nm (diffusion coefficient of 0.2-12×10$^{-3}$ cm$^2$/s) for chalcogen-based quantum dot

(QD) assemblies,[7, 9, 13] and of 3-50 nm for organic systems where the exciton transport occurs *via* singlet emitters and therefore FRET.[14-17] The short diffusion lengths are attributed to one or several of these parameters to be limiting, for instance the spatial disorder in QD assemblies.

PNCs are a fascinating NC class with the potential to excel in all of these parameters. As bulk semiconductors, lead halide perovskites have emerged as a promising class of materials for low-cost, solution-processable optoelectronics,[18-21] demonstrating thin-film solar cell power conversion efficiencies exceeding 25%[22,23] and LEDs with 20% external quantum efficiency.[24-27] In the form of NCs, the all inorganic cesium lead halide ($CsPbX_3$, X = I, Br, or Cl) PNCs provide high optical tunability as a function of composition,[28-31] size,[30,32-35] and shape,[32,33,36] with impressive exciton generation efficiency,[37] and scalable solution-phase processes.[38-41] Although the quantitative use of the FRET equation is beyond the scope of this work, we note that in terms of FRET-mediated transport they qualitatively optimize all 6 parameters: (1) The inter-particle distance $r$ is determined by the ligand (oleic acid and oleylamine) and amounts to 2 nm. (2) The spectral overlap is high due to narrow emission linewidths and small Stokes shifts.[28] Furthermore, their defect-tolerant optical emission (responsible for record efficiencies in QD solar cells)[29] maintains the symmetric and narrow emission bands compared to chalcogen based NCs,[21] and hence the high spectral overlap. The transition dipoles (3) between adjacent PNCs have been reported to be well-aligned as well,[42] due to their cubic shape directing the overall orientation in a self-assembled layer. PNCs have also demonstrated a combination of high photon absorption crosssection (4) and near unity quantum yield (5).[43] A flat energy landscape (6) among NCs can be obtained by controlling the NC size in a weakly quantum confined regime where the energy bandgap is primarily determined by the chemical composition, so as to reduce the thermalization of excitons towards NCs with smaller bandgap, which would limit the back transfer onto larger bandgap NCs and therefore decrease the sites availability for the exciton random hopping.[28]

In this work we show that indeed, FRET-mediated exciton transport through self-assembled close-packed PNC mono layers reaches over 200 nm with a diffusivity of ~ 0.1 cm$^2$/s, which is in terms of diffusion length one order of magnitude higher and in terms of diffusivity two orders of magnitude larger than previously reported for chalcogen based QDs.[9] We demonstrated these values by studying the exciton diffusion in a close-packed monolayer of CsPbBr$_3$ PNCs. Importantly, in order to create a model system for the study of such diffusion, we took particular care in the sample preparation and developed a strategy to deposit only one monolayer of PNCs per sample. The ordered self-assembly of PNCs in a monolayer causes the exciton propagation to be restricted to two dimensions only, allowing us to directly image the exciton motion on a 2D plane without any loss of data in the third dimension. Moreover, this 2D PNCs assembly can be replicated theoretically with a 2D random walk, thus simplifying the interpretation of experimental data within a statistic framework. The diffusion length and the diffusivity were determined *via* stead state PL microscopy mapping and the diffusivity using time resolved PL mapping, respectively. The diffusion was modeled using continuum and discrete representations of exciton hopping to provide a physical interpretation of our experiments, showing that albeit demonstrating record FRET mediated exciton diffusion length, our system is still in a sub-diffusive regime.

DISCUSSION

To simplify the direct visualization of exciton transport we confined the PNCs in a self-assembled monolayer on a Si wafer with about 1 nm of native oxide and coated with ~10 nm of a hydrocarbon polymer, deposited *via* plasma-polymerization of methane, to prevent both exciton quenching into the Si and wave-guiding along the dielectric layer. We made two kinds of samples, a close-packed sample to study the exciton diffusion, and a control sample with a sparse layer of small groups of PNCs separated by at least 20 nm to prevent FRET based diffusion. The NCs examined in this study were cubic $CsPbBr_3$ PNCs with an average side length of 10 nm, synthesized according to an established procedure.[28] Spin-coating from a toluene solution on a surface functionalized with a –CH terminated polymer reproducibly yielded close-packed monolayers with an inter-particle spacing of ~2 nm, which is determined by the ligand length (see Fig. 1, Methods, and the SI for a detailed description of the sample preparation methods). In order to stabilize the PNCs during optical characterization (conducted at room temperature and under ambient atmosphere), we adapted a recently reported passivation process:[44] after spin-coating the solution of PNCs, we deposited ~3 nm of aluminum oxide by plasma-assisted atomic layer deposition (ALD). The passivated samples emitted stable PL upon illumination with continuous wave (CW) laser powers as high as 2000 $W/cm^2$ or with pulsed laser fluences as high as 50 $\mu J/cm^2$.

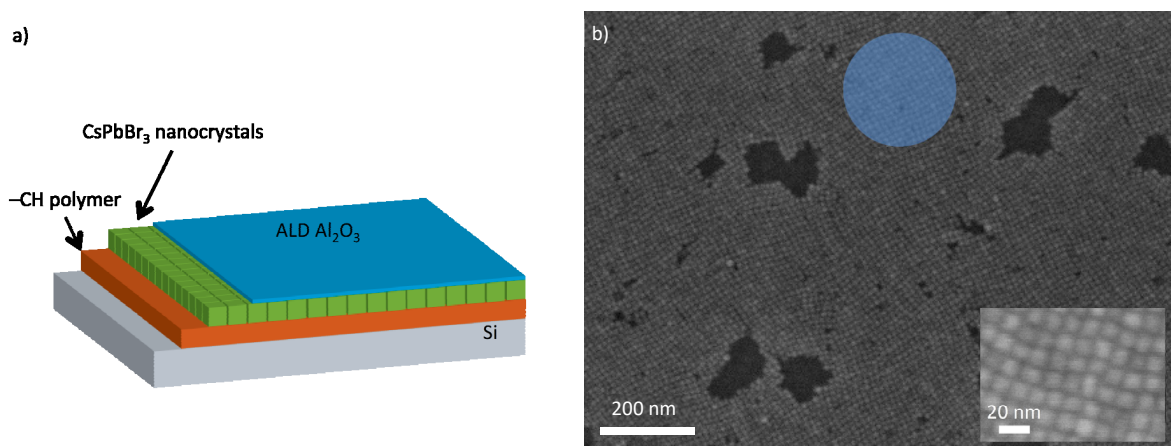

**Figure 1. Deposition of controlled PNC 2D architectures.** (a) CsPbBr$_3$ PNCs in toluene were spin-coated onto a Si substrate functionalized with a –CH terminated polymer and coated with 3 nm of aluminum oxide deposited by ALD to prevent degradation during measurements. (b) SEM micrograph of a close-packed monolayer of PNCs. The blue circle shows the size (FWHM ~200nm) of the excitation laser spot used in subsequent optical experiments.

Exciton diffusion in NC solids and organic semiconductors has predominantly been studied indirectly by spectroscopic techniques, which can only provide a coarse estimate of the diffusion length.[7, 45] Recent microscopy-based approaches allow for the direct measurement of diffusion dynamics including the two main quantities that characterize exciton diffusion: the diffusion length and the diffusion coefficient (or diffusivity). [9]

Figure 2 illustrates the basic principles of the visualization of exciton diffusion. In order to measure the diffusion length and coefficient, we start by exposing the PNCs monolayer to a perpendicular laser beam, whose size is kept as close as possible to the diffraction limit. The laser beam generates a local population of excited states in the PNCs monolayer, with an initial spatial distribution that matches the intensity profile of the excitation spot. The PL maps were recorded with a CCD camera, after sample excitation with a 450nm laser, whose spot was kept as close as possible to the diffraction limit, with a FWHM of 240nm (Supplementary Fig. S18). If the motion of an exciton is allowed, for instance by FRET-induced hopping onto a nearby PNC, then it is possible that such exciton will travel before radiatively recombining. Our sample preparation greatly enhances the probability of such hopping by promoting the self-assembly of the PNCs in a close-packed monolayer. Excitons can therefore propagate within the 2D plane of the monolayer, which ultimately results in a radial expansion from the excited state distribution (Figure 2a) and in a broader spot of the collected PL with respect to the excitation spot. Conversely, if the distance increases with respect to the close-packed case, the FRET-hopping is not allowed, and the emission can only occur from the same PNC that was initially excited (Figure2b). In order to better understand this effect, we report in Figure 2b-c two images of PNCs, one in a close-packed lattice and the other one when the PNCs are spincoated onto a non-engineered substrate, thus resulting in a random distribution where each PNCs is far enough from others (at least 20nm) to hinder FRET transport. The collected PL from such samples are reported in Figure 2e,f. Here, we can appreciate the significantly different results of the excitation with the same spot, shape and energy: the spatial extent of the PL map collected from the close-packed sample is much broader than the excitation spot, whereas in the sample in which excitons cannot hop the spatial extent of the PL map resembles the excitation spot, only slightly enlarged due to the convolution of the point-spread function of the microscope with the spatial profile of the excitation spot PNC (*i.e.*, a convolution of a diffraction-limited point source with the diffraction-limited spot of a focused laser beam; see Supplementary Fig. S18). A schematic of the set-up is shown in Supplementary Fig. S19. In order to provide a more quantitative comparison between the two images, in Figure 2g we report a line scan through both PL maps as extracted from Figures 2e-f. Both spots are radially symmetric, so we can take any direction in each spot and compare their relative cross-sections. In first order approximation it is possible to describe the diffusion profile within a Gaussian framework. By subtracting the variances of the Gaussian fits with and without diffusion (close-packed *vs* sparse monolayer), we obtain a first estimate of the diffusion length, as explained in detail in the Supporting Information (see section S2). In order to provide a theoretical support for our data, we perform a detailed set of calculations, which are reported in the Supporting Information (see Sections S3 to S6). Specifically, we model the steady state diffusion using two complementary approaches, namely: *i)* continuum equations (S4) and *ii)* microscopic simulations (S5). First, we approximate the processes of creation, hopping, and recombination of optically excited excitons in nanoparticles by considering the statistic of classical bosonic particles on a lattice, in steady-state conditions. This description results in the derivation of a continuity equation, where the concentration of excitons is the physical quantity that undergoes a spatial variation with its own diffusion constant. The equation is solved both numerically and semi-analytically, with agreement between the two solutions in the linear regime, *i.e.*, the condition of low excitation power in which the optical measurements were conducted. The semi-analytic solution is particularly useful to simulate the trends in non-linear regimes, that is, when the excitation power is large enough to prompt a power-induced change of the exciton diffusion profile. The modelling *via* continuity equation ultimately indicates that our PL profiles are consistent with a mean excitonic diffusion constant of the order of 0.5 cm$^2$/s, or (224 nm)$^2$/ns, which results in a characteristic exciton diffusion length of about 200 nm (consistent with the simple estimate in which we approximate profiles as Gaussian distributions).

In addition, we perform a set of discrete-time and continuous-time Monte Carlo simulations, where we model excitons as classical particles, able to undergo various processes, sitting with fixed positions within a 2D square array, similarly to the experimental sample. Using a model that assumes a completely uniform film of NCs, subtle but significant variations occur between the tails of measured and predicted profiles. However, as seen in the SEM images, vacancies occur in the experimental film (Fig. 1b). These vacancies can inhibit further exciton propagation. Hence, in order to provide a more realistic description of our system, we carry out Monte Carlo simulations in presence of both spatial and energetic disorder. Spatial disorder is taken into account by arranging the available site in a non-close-packed fashion, so that the hopping of an exciton onto a neighbour PNC may be inhibited due to excessive distance from the closest neighbour, whereas energetic disorder takes into account the deviation from perfectly isoenergetic particles and hence the presence of larger or smaller PNCs, resulting in (slightly) smaller or larger bandgap energies. When such disordered conditions are included in the model, the predicted PL profiles reproduces the experimental PL profile more closely (Fig. 2g).

Importantly, the excitation power in all measurements was kept sufficiently low so as to remain in the linear excitation regime (see Methods and Supplementary Fig. S23). The presented interpretation of deconvolving the Gaussian distributions and extracting the diffusion lengths is only correct in the regime where PL intensity scales linearly with the excitation power. At higher pump fluences, exciton-exciton annihilation starts taking place, thus introducing an additional term in the diffusion equations that may lead to significantly broader PL profiles that could be interpreted as enhanced exciton diffusion, while actually it is related to a steep exciton concentration gradient and the effective diffusion length for each exciton is still the same. A detailed discussion is reported in the Supporting Information.

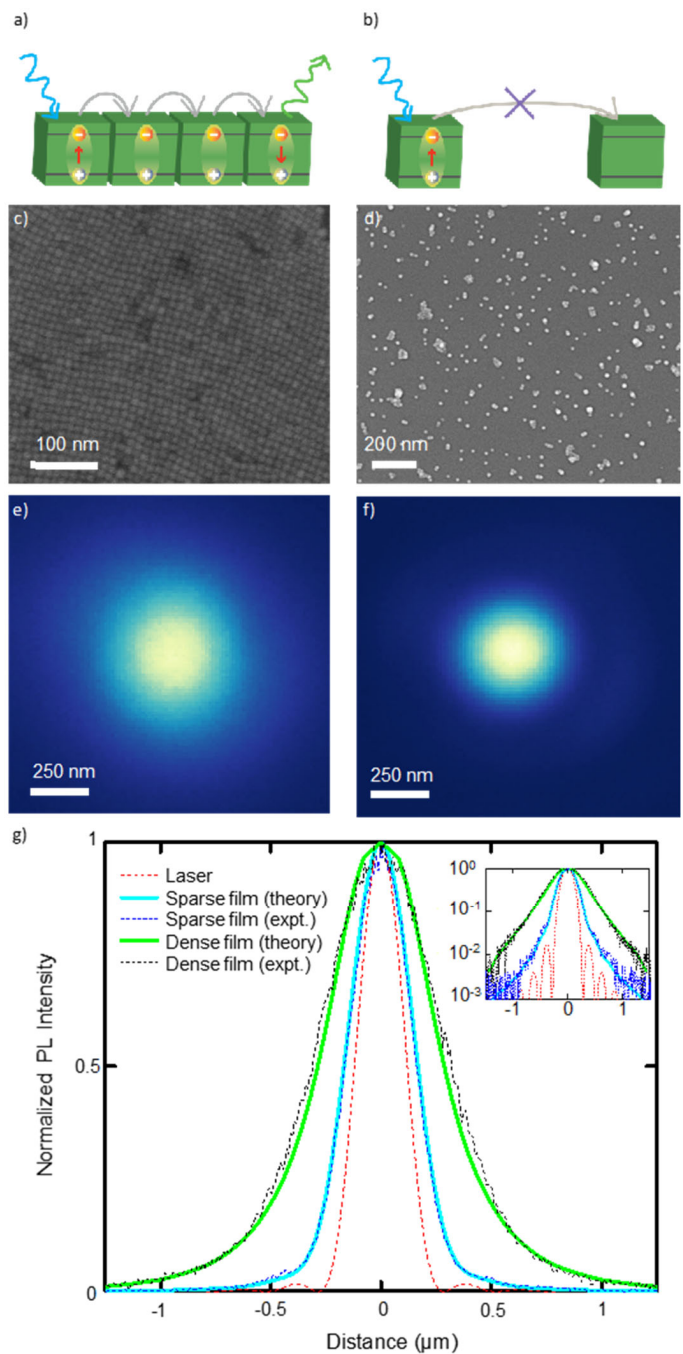

**Figure 2. Direct measurement of steady-state exciton diffusion.** (a) When PNCs are assembled in a close-packed monolayer, the distance between NCs is minimized, allowing for efficient FRET-mediated exciton diffusion. (b) When PNCs are spatially separated, FRET-mediated exciton diffusion is inhibited. (c) SEM micrograph of a close-packed monolayer of PNCs. (d) SEM micrograph of a sparse monolayer of PNCs. (e) Normalized PL intensity profile emitted by the close-packed PNCs monolayer when excited with a diffraction-limited laser spot with wavelength 450 nm. (f) Normalized PL intensity profile emitted by the sparse PNCs monolayer when excited with a diffraction-limited laser spot with wavelength 450 nm. (g) PL profile cross-sections of panel (e) (black dashed line) and (f) (blue dashed line) together with simulated PL profile cross-sections for a square lattice of nanoparticles with a vacancy fraction of 20% (green line), and for a sparse sample of nanoparticles on which hopping cannot occur (cyan line). The dashed red line corresponds to the excitation laser profile cross-section. The inset shows the main figure on a logarithmic vertical scale.

CsPbBr$_3$ PNCs are not strongly quantum confined in the size range used in this study (10 nm average cube side);[28] as a result, our close-packed monolayers mostly constitute a flat energy landscape for exciton diffusion, which is a requirement to maximize FRET in a NC solid. In the presence of energetic heterogeneity, excitons travel downhill in energy and thermalize onto NCs with smaller bandgaps. The process of back transfer onto larger bandgap energies has a lower rate and is less likely to take place, thus limiting the availability of viable neighbours for an exciton to hop on. Because of this more efficient funneling of excitons from high-energy to low-energy states than vice-versa, signatures of this inhomogeneity-driven process can be discerned in the temporal evolution of the PL spectrum after pulsed excitation. The additional energy transfer relaxation channels of the high-energy sites accelerate their relaxation rates, yielding energy-dependent excited-state lifetimes, where higher-energy NCs have shorter lifetimes than those with lower-energies. Thus, following pulsed excitation, the PL spectrum typically shifts to lower energies as the higher-energy NCs relax more rapidly than the low-energy NCs. Experimentally, this behavior can be resolved as time-dependent shifts in the PL spectrum to lower energy or emission energy-dependent relaxation kinetics. In colloidal semiconducting NCs, this energetic heterogeneity largely arises from the polydispersity of the sample where the larger quantum dots are less quantum confined, have lower-energy excited states, and thus serve as sites where an exciton can easily transfer to, but from where it cannot easily leave.[46,47]

In Figure 3, we used time-resolved PL (TRPL) spectroscopy to assess the disorder in the energetic landscape in our films. As shown in Fig. 3a, during the assembly process, energy minimization pushes the PNCs of different size to the edges of the ordered, close-packed regions. The result is that the central regions of the film are significantly less polydisperse than the edges. Examples of a uniform area (top dashed rectangle) and of a non-uniform area (bottom dashed rectangle) are highlighted in Fig. 3a. As seen in Figs. 1b and 2c, the exciton diffusion studies were performed in the regions with monodisperse PNCs; in these portions of the film time-resolved PL spectroscopy measurements displayed identical decay kinetics for the high and low energy components of the PL spectrum (Fig. 3b) and, accordingly, the PL spectrum does not change with time (Supplementary Fig. S20). In contrast, in the edge regions with maximal polydispersity, signatures of excitons getting funneled in lower-energy bandgap NCs were subtle but detectable: the lower energy emission (2 nm band; centered at 529.2 nm) decays slower than the higher-energy emission (2 nm band; centered at 502.6 nm) and the PL spectrum exhibits a clear evolution to lower energies following pulsed excitation (Supplementary Fig. S20). These results indicate that within the central regions of the film, the site-to-site variations in energy are minimized thus strongly reducing, or possibly eliminating, exciton funneling towards low-energy sites. Therefore, only at the edges, where the polydispersity is maximal, we see evidence of an energetic landscape where the energy transfer towards smaller bandgap NCs is more efficient. Rather, in the central regions, the weak quantum confinement combined with the assembly process play a pivotal role in achieving the long exciton diffusion distance. Future work will be dedicated to a detailed investigation of the redshift observed in defectual areas.

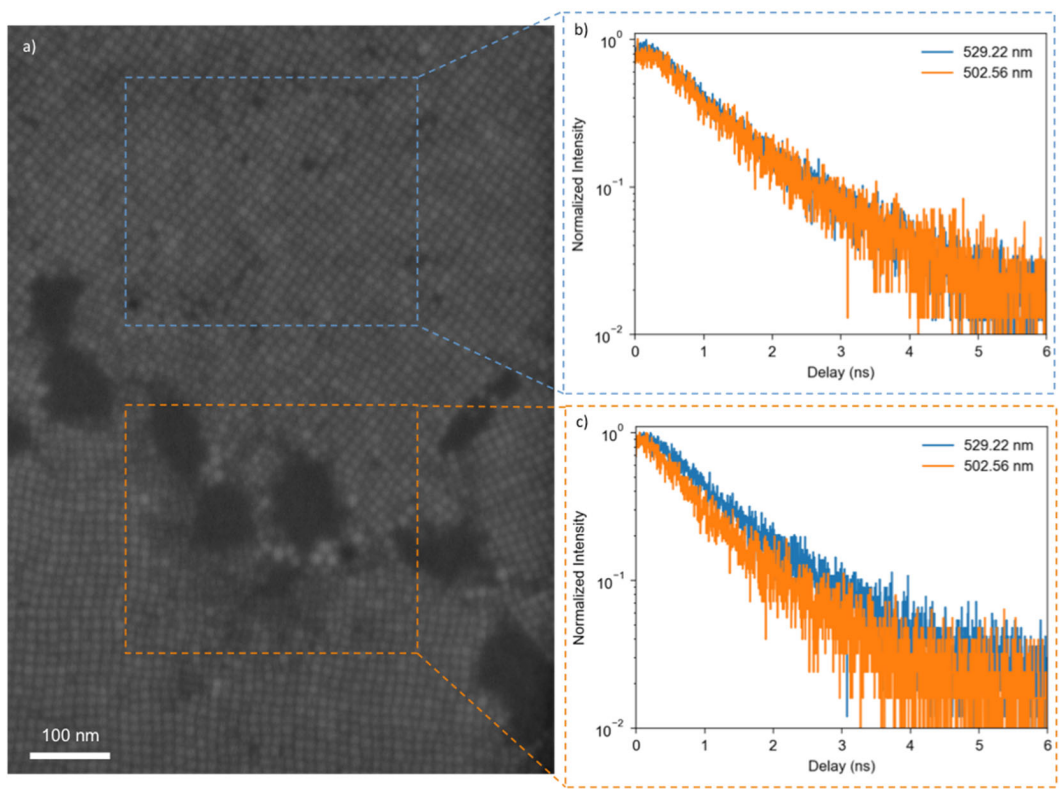

**Figure 3. Probing the energy landscape by time-resolved PL spectroscopy.** (a) SEM micrograph of a close-packed monolayer of PNCs showing ordered areas made of uniformly sized

PNCs (blue dashed rectangle) and disordered areas made of PNCs of different sizes at a crack in the film (orange dashed rectangle). (b) PL intensity as a function of time at two emission wavelengths, 529.22 nm (blue solid line) and 500.56 nm (orange solid line), measured on an ordered area made of uniformly sized PNCs. The overlap of the curves indicates equivalent PL lifetimes at both emission wavelengths. (c) PL intensity as a function of time at two emission wavelengths, 529.22 nm (blue solid line) and 500.56 nm (orange solid line), measured on a disordered area made of PNCs of different sizes. The orange curve displays the slower decay of the low energy portion of the PL spectrum, due to exciton migration to smaller bandgap PNCs.

To better understand and measure the dynamics of the long-distance exciton diffusion, time-resolved PL microscopy was employed to track the temporal evolution of the PL spatial profile as it expanded from its initial state. Areas of the close-packed monolayer of monodisperse PNCs were excited with short laser pulses and the magnified PL emission (100×) was collected by a single mode fiber (with a diameter of 5 μm). To gain spatial resolution, the collection fiber was mounted on a translation stage that systematically scanned the fiber aperture in the focal plane of the microscope. At each fiber position, a full time-resolved PL transient was recorded, which allowed us to track how the excited state expanded with time. The temporal evolution of the normalized PL profile is shown in Figure 4a, where the 0 nm distance corresponds to the position of the fiber, which is aligned with the center of the PL spot. Distances greater than zero are reported as the distance between the center of the PL spot and the center of the image of the fiber aperture on the sample plane. The spatially and time-resolved PL trace in Figure 4a clearly displays the expected diffusive broadening, which can be quantified by calculating the PL profile width at each time slice (Fig. 4b). The rate at which the variance of the spatial PL profile (Gaussian fit) rises with time is sub-linear, indicating a sub-diffusive process, which, as discussed above, is very likely due to PNC voids in the film. In these dynamic measurements, as reported in previous work,[9] the average exciton diffusion length is calculated as the increase in the PL profile variance from its initial state (right after excitation) to the average lifetime of the system. The average lifetime of these PNCs in a close-packed film is 1.14 ns (as calculated as the time to reach a decay of 37% or 1/e of the PL intensity, see Supplementary Fig. S21), which, using the dynamics in Figure 4b, yields an average exciton diffusion length of 194 nm that is in very good agreement with that derived from our steady-state measurements.

The PL lifetime was also calculated at each point (a PL lifetime map is shown in Supplementary Fig. S22) and was observed to decrease as the reciprocal of the PL intensity (Fig. 4c). This spatially dependent change in dynamics is an expected effect of the diffusion process, which is driven by the exciton density gradient, and results in a larger net outward exciton flux in areas with a large exciton population. The minimum lifetime was found to be 1 ns, and grew to 1.3 ns 500 nm away from the excitation intensity maximum. The majority of the signal (>90%) lies between -500nm and 500nm from the excitation maximum; as a result, the lifetimes calculated for values <-750nm and >750nm result very scattered, since the diffusion outside this region is almost completely exhausted and the signal is weak and indistinguishable from noise (especially considering that we are collecting with a single mode fiber with a diameter of 5 μm). Importantly, we highlight that the measurements were performed with excitation intensities that are deep in the linear regime of the power-dependence of the PL (see Supplementary Fig. S23), therefore we exclude the possibility that the central decrease in lifetime is due to higher order non-radiative recombination processes (*e.g.* Auger recombination).

The diffusion coefficient (or diffusivity) was calculated as the first derivative of the PL profile variance increase in time and is shown in Figure 4d. The effective diffusivity was found to decrease with time, from 0.5 cm$^2$/s right after excitation to ~ 0.1 cm$^2$/s at the end of the diffusion process.

These values of FRET-mediated exciton diffusion length and diffusivity are the largest reported so far for NC solids. Chalcogen-based QDs demonstrated diffusion lengths in the range of 20-30 nm and diffusivities between 0.2E-3 cm$^2$/s and 1.5E-3 cm$^2$/s.[9] The system here reported demonstrated one order of magnitude higher diffusion length and about two orders of magnitude higher diffusivity. Significantly, the exciton diffusion dynamics in our PNC films is comparable with that of unbound charge carriers in other bulk perovskite materials. Individual crystals of CsPbI$_2$Br perovskite measured by pump-probe microscopy revealed a diffusivity of 0.27 cm$^2$/s,[48] and thin films of polycrystalline hybrid perovskite measured by transient absorption microscopy showed a diffusivity in the range of 0.05 cm$^2$/s to 0.08 cm$^2$/s, with diffusion lengths of 220 nm in the 2 ns experiment time.[49]

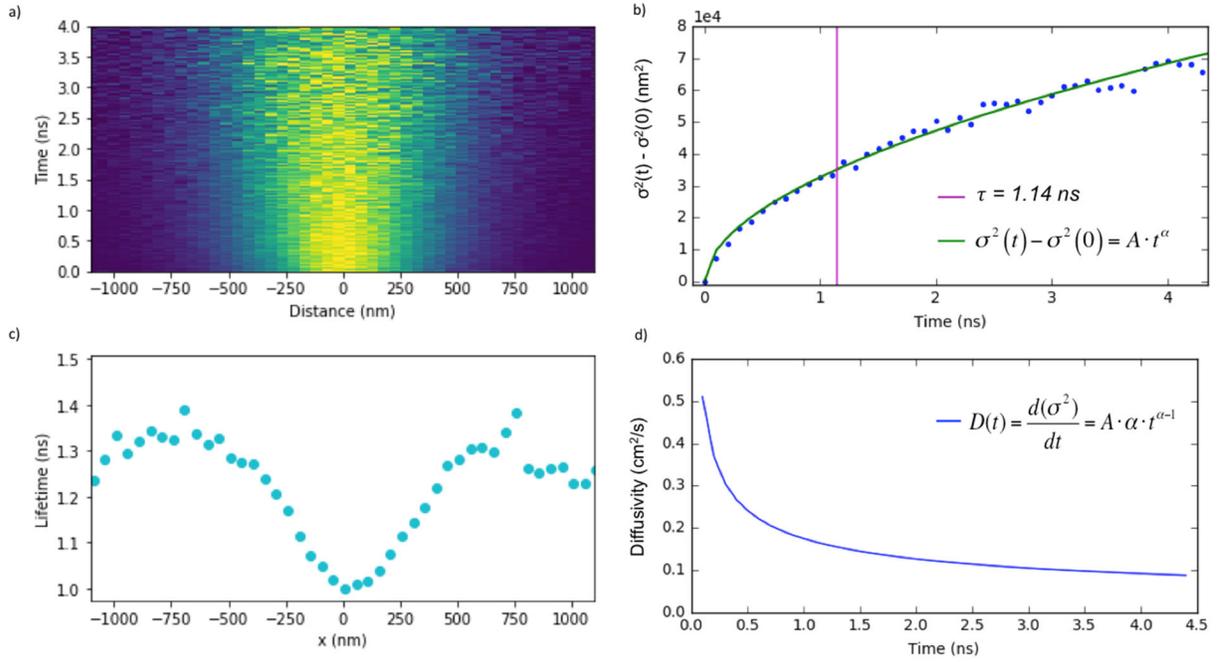

**Figure 4. Probing exciton diffusion dynamics by time-resolved optical microscopy.** (a) Time evolution of cross-sectional PL intensity profile. (b) PL profile variance increase as a function of time (blue dots). The green solid line shows the fit to the power law A·t$^\alpha$ (A = 0.3527 cm$^2$/s , $\alpha$ = 0.53). (c) Space-resolved PL lifetime calculated from the signal in (a) as the time for 37% or *1/e* decay of PL intensity. (d) Diffusivity (or diffusion coefficient) as a function of time calculated as the first derivative of the PL profile variance variation in (c).

CONCLUSIONS

In conclusion, ordered monolayers of isoenergetic PNCs were fabricated *via* controlled self-assembly. This system demonstrated extremely efficient FRET-mediated exciton diffusion, which was directly characterized by steady-state and time-resolved PL microscopy together with an analytical and statistical model that granted a deeper understanding of the exciton diffusion dynamics. Our measurements directly capture a diffusion length of 200 nm and a diffusivity of 0.5 cm$^2$/s. This diffusion length is ten times longer and the diffusivity is two orders of magnitude larger than previously reported values for films of chalcogen-based QDs,[9,50] and comparable to that of charge-carrier diffusion in thin films of polycrystalline hybrid perovskite. Demonstrating such long-range diffusion, our PNC system is ideally suited to study FRET processes and FRET-mediated energy transfer on length scales that are easily accessible and therefore easy to optimize. Moreover, long exciton diffusion establishes an additional design element for next generation PNC-based optoelectronic devices.[21, 51]

Further progress towards even longer-range exciton diffusion may be achieved by improving PNCs energetic uniformity as well as by optimizing the protective ALD-based process that prevents perovskite degradation. Additionally, fabricating PNC assemblies with increased complexity, for example deliberately varying inter-particle distance in certain assembly portions or by positioning PNCs with decreasing bandgap next to each other forming an oriented energy funnel, could demonstrate ways to move excitons to predetermined positions. Overall, we showed that ordered assemblies of isoenergetic PNCs support FRET-mediated exciton diffusion with exceptional lengths, which can be used to better the performances of PNCs-based optoelectronic devices.

METHODS

*PNCs synthesis and characterization*

CsPbBr$_3$ nanocubes were synthesized by a procedure adapted from the original report.[28] All chemicals were purchased from Sigma-Aldrich and used as received without further purification. Cs$_2$CO$_3$ (1.2 mmol) was added to 10 mL 1-octadecene and stirred at 120 °C under vacuum for 1 hour. Oleic Acid (2 mmol) was injected under nitrogen atmosphere and resulting mixture was stirred at 120 °C for two hours until fully dissolved. In a separate container, PbBr$_2$ (0.19 mmol) was added to 5 mL 1-octadecene and stirred at 120 °C under vacuum for 1 hour. Oleic acid (1.6 mmol) and oleylamine (1.5 mmol) were injected under nitrogen atmosphere. The resulting mixture was stirred at 120 °C for two hours until fully dissolved, then was heated to 165 °C. To this preheated lead solution was added 0.4 mL of hot Cs$_2$CO$_3$ solution under nitrogen atmosphere with vigorous stirring. Reaction was stirred for 5 seconds and cooled rapidly in ice bath until reaction mixture solidified. After freezing, reaction mixture was warmed to room temperature and transferred into centrifuge tubes. The mixture was centrifuged at 8,500 rpm, for 10 minutes. The supernatant was discarded, and the pellet was redispersed in anhydrous hexane (6 mL). An equal volume of tert-butanol was added to precipitate the NCs, and the mixture was centrifuged at 12,000 rpm for 15 minutes. The supernatant was discarded, and the pellet was redispersed in toluene. These solutions were then centrifuged for 5 minutes at 700 rpm, and the pellet was discarded to remove large aggregates. The supernatant was transferred to a glove box for film deposition.

*PNCs monolayer fabrication*

PNCs monolayers were prepared by spin coating (1,500 rpm, 45 s) from a colloidal suspension of nanocubes in toluene onto Si wafers coated with 10 nm of a –CH terminated polymer, which is sufficiently thin to prevent lateral wave-guiding but thick enough to prevent exciton quenching by the silicon. The concentration was 3 g/l for the close-packed monolayer and 60 mg/l for the sparse monolayer. The hydrocarbon polymer was deposited by polymerizing methane in a plasma chamber (40 mTorr, RF power 100 W, 10 °C, Oxford Instruments). Aluminum oxide (3 nm) was deposited by plasma assisted atomic layer deposition at 40 °C (Oxford Instruments). Films were characterized by SEM (Zeiss).

*Steady-state PL microscopy*

The setup for steady-state PL microscopy is shown in Supplementary Figure S3a. The 450 nm CW diode laser source was collimated and then focused to a diffraction-limited spot by a 100X 0.95 NA objective lens. The back aperture of the objective was overfilled to assure diffraction-limited performance. Emission from the sample was collected by the same objective and additionally magnified 5.3X for a total magnification of 530X and imaged on a CCD camera (QSI SI 660 6.1mp Cooled CCD Camera) with pixel size 4.54 μm, which provided an effective imaging pixel size of 8.63 nm. A 490 nm long-pass dichroic filter (Semrock Di03-R488-t1) and two 496 nm long-pass edge filters (Semrock) were used to remove the excitation laser beam from the PL signal. The laser beam was imaged through the 490 nm long-pass dichroic filter (Semrock) and a 498 nm short-pass edge filter (Semrock) to remove the PL signal. Measurements were performed at 45 W/cm$^2$, which corresponds to the probability of one absorbed photon per 1,705 nanocubes during the 1.15 ns average lifetime.

*Time-resolved PL spectroscopy*

The setup for time-resolved PL microscopy is shown in Supplementary Figure S3b. The pulsed laser source (center wavelength 465 nm with a 2.5 nm bandwidth; 5 ps pulse duration; 40 MHz repetition rate) was collimated and focused by a 100X 0.95 NA objective lens. The back aperture of the objective was overfilled to assure diffraction-limited performance. Emission from the sample was collected by the same objective. A 490 nm long-pass dichroic filter (Semrock) and a 496 nm long-pass edge filter (Semrock) were used to remove the excitation laser beam from the PL signal. The PL spectral components were separated with a monochromator (Princeton Instruments Acton 2300i) and detected by a single-photon counting avalanche photodiode (MPD PDM series) connected to a time-correlated single-photon counting unit (Picoharp 300). The temporal resolution was approximately 50 ps as determined by the FWHM of the instrument response function.

Measurements were performed at laser fluence 2.5 µJ/ cm$^2$, which corresponds to the average probability of one absorbed photon per 32 nanocubes per pulse.

*Time-resolved PL microscopy*

The setup for time-resolved PL microscopy is shown in Supplementary Figure S3c. The pulsed laser source (center wavelength 465 nm with a 2.5 nm bandwidth; 5 ps pulse duration; 40 MHz repetition rate) was collimated and focused by a 100X 0.95 NA objective lens. The back aperture of the objective was overfilled to assure diffraction-limited performance. Emission from the sample was collected by the same objective and imaged on a single-mode fiber (P1-405P-FC-2, Thorlabs) attached to a translation stage (Attocube ECS series) that scanned the emission focal plane. The fiber mode field diameter was 2.5 – 3.4 μm at 480 nm; the stage was moved in 5 μm steps corresponding to 50 nm at the sample. A 490 nm long-pass dichroic filter (Semrock) and two 496 nm long-pass edge filters (Semrock) were used to remove the excitation laser beam from the PL signal. The laser beam was imaged through the 490 nm long-pass dichroic filter (Semrock) and a 498 nm short-pass edge filter (Semrock) to remove the PL signal. The signal was detected by a single-photon counting avalanche photodiode (MPD PDM-series) connected to a time-correlated single-photon counting unit (Picoharp 300). The temporal resolution was approximately 50 ps, as determined by the FWHM of the instrument response function.

Measurements were performed at laser fluence 5 μJ/ cm$^2$, which corresponds to the average probability of one absorbed photon per 16 nanocubes per pulse.

The sample was mounted above the objective lens on a piezoelectric scanning stage. Samples were scanned during the course of the measurements (~30 min) over an area of 5×5 μm to avoid photobleaching or photodamage.

*Physical modeling*

We simulated the processes of exciton creation, recombination, and hopping at continuum- and microscopic levels of resolution. We convolved the resulting excitonic profiles with the optical point-spread function in order to calculate observed PL profiles. Full details of these calculations are given in SI sections S3 to S6.

For Table of Contents Only

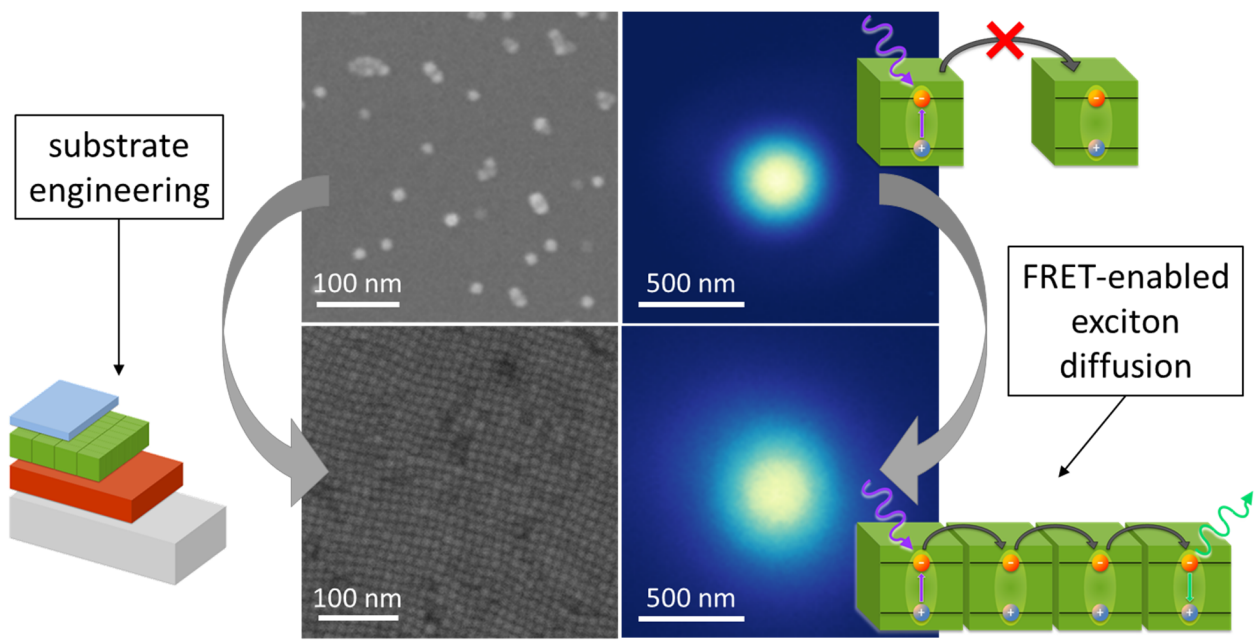

substrate engineering

FRET-enabled exciton diffusion

ACKNOWLEDGMENT

This work was performed at the Molecular Foundry supported by the Office of Science, Office of Basic Energy Sciences, of the U.S. Department of Energy under Contract No. DE-AC02-05CH11231. M.J.J. and Y.L. were also supported by the U.S. Department of Energy, Office of Science, Office of Basic Energy Sciences, Materials Sciences and Engineering Division, under Contract No. DE-AC02-05CH11231 within the Inorganic/Organic Nanocomposites Program (KC3104). A.W.-B. was supported by the U.S. Department of Energy Early Career Award.

We thank Prof. Alexander Holleitner (Technical University Munich) and Prof. Ian Sharp (Technical University Munich) for insightful discussions.

ASSOCIATED CONTENT

**Supporting Information**. Supporting Information Available: the file contains detailed description of the sample preparation and of the estimation of the diffusion length with a Gaussian approximation, a comprehensive section of the theoretical modeling *via* both continuity equations and Monte Carlo simulations, and Supporting Figures. (PDF). This material is available free of charge *via* the Internet at http://pubs.acs.org.

A preprint version of this manuscript is deposited in arXiv:

## AUTHOR INFORMATION

**Present Addresses**

† Department of Physics, Montana State University, PO Box 173840, Bozeman, MT, 59717

**Author Contributions**

The manuscript was written through contributions of all authors. R.B. and A.W.-B. conceived the initial work. A.L. and I.R. performed initial experiments. E.P. prepared the samples, conducted the diffusion experiments and analysis, and wrote the manuscript. S.W. developed theoretical modeling and analysis. A.L. and M.J.J. synthesized the nanocrystals with supervision of R.B. and Y.L. E.S.B. and N.J.B. supported the experimental measurements and data analysis. E.P., E.K.W., and E.S.B. built the microscope. A.M.S. and S.C. supported the sample preparation and provided input into the data interpretation. E.P., E.S.B., N.J.B., M.J.J., A.M.S., S.W., M. L. and A.W.-B. helped with writing the manuscript. All authors have given approval to the final version of the manuscript.

# Long-Range Exciton Diffusion in Two-Dimensional Assemblies of Cesium Lead Bromide Perovskite Nanocrystals

Erika Penzo, Anna Loiudice, Edward S. Barnard, Nicholas J. Borys, Matthew J. Jurow,
Monica Lorenzon, Igor Rajzbaum, Edward K. Wong, Yi Liu, Adam M. Schwartzberg,
Stefano Cabrini, Stephen Whitelam, Raffaella Buonsanti, and Alexander Weber-Bargioni

## S1. PEROVSKITE NANOCRYSTAL (PNC) SAMPLE PREPARATION FOR EXCITON DIFFUSION MEASUREMENTS BY PL MICROSCOPY

Exciton diffusion can be controlled by modulating the NC assembly. Minimizing NC-NC distance (R) is essential to maximizing the rate of FRET (FRET rate is inversely proportional to $R^6$). In a 3D NC solid, excitons can move in any direction within the solid, as long as the NCs are physically close enough and with sufficient spectral overlap. Confining the NCs to 2D reduces the available paths for exciton hopping to a space directly accessible to imaging by optical microscopy.

The solution concentration and the spinning parameters were optimized for close-packed monolayer deposition; all samples were deposited at 1,500 rpm for 45 seconds and the film morphology was adjusted by varying the concentration of PNCs solution. A closed-packed monolayer without portions of an extra layer and with minimum empty regions was consistently obtained when spin-coating from a *ca.* 3 g/l solution in toluene of CsPbBr$_3$ PNCs at 1,500 rpm for 45 seconds. A sparse monolayer was obtained when spin-coating from a *ca.* 60 mg/l solution of CsPbBr$_3$ PNCs at 1,500 rpm for 45 seconds. A closed-packed monolayer of NCs was not achievable without the surface functionalization with -CH ter-minated polymer (Supplemental Fig. S1) and/or with solutions in solvents other than toluene. Hexane and octane solutions deposited only patches of multilayers separated by wide empty regions. Development of a reproducible and controllable deposition technique was necessary for establishing consistent PNC monolayers. Simple drop casting, due to the lack of control over the drying process, yielded unacceptable sample-to-sample variability in the final film morphology. Although very effective in forming highly ordered 2D assemblies of conventional semiconducting QDs, methods depending on the interface between immiscible solvents (*e.g.* Langmuir-Blodgett techniques) were not applicable to PNCs because of their instability (solubility) in polar solvents[1]. Spin-coating demonstrated reproducible results and allowed control of the density of PNCs in the film by modulating either spin-coating speed or solution concentration. The morphology of the film was mostly determined by functionalization of the substrate surface and by the solvent in which the PNCs were dispersed. We achieved the best control of film morphology when the substrate surface was functionalized with a -CH terminated polymer and the PNCs were dispersed in toluene. Under these conditions, spin-coating from a dilute solution would yield separated patches of PNCs, with patch sizes controllable all the way to individual PNCs. The distance between individual patches increased as the PNC concentration was reduced. Conversely, deposition from a concentrated solution yielded a continuous monolayer of PNCs, sporadically covered by patches of a second layer. The size of the patches increased, and the distance between them decreased, with increasing solution concentration. For very concentrated PNC solutions, two complete monolayers were formed. The number of layers steadily increases with increasing PNCs concentration in solution, following the same process of patch expansion. The spin-coating speed had a similar effect on film morphology. High spin-coating speeds would reduce the film thickness or increase the spacing between NCs patches and reduce the patches' size, while low spin-coating speed would have the opposite effect. Adjusting the spin-coating speed had a more limited impact on the film morphology. Adjusting the NP concentration was the most effective way to control the PNCs film density. All samples used in this study were screened by SEM to ensure consistent morphologies and to compensate for batch to bath variability common to PNCs.

Alumina coating by ALD was necessary to ensure sam-ple stability under illumination and to prevent long-term degradation due to air moisture. Unprotected PNCs samples were found to be unstable under illumination with a focused laser beam even at low excitation power (below 100 W/cm$^2$). The main signatures of this instability were a rapid decay of photoluminescence (PL) intensity together with changes in the PNCs film morphology, observed by SEM after exposure to the laser beam.

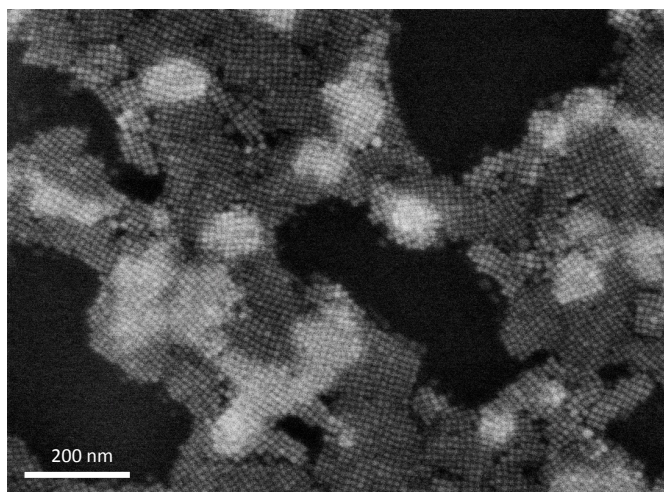

FIG. S1: SEM micrograph of PNCs deposited from a toluene solution (concentration ca. 3 g/l) by spin-coating (1,500 rpm for 45 s) on a Si wafer.

## S2. ESTIMATION OF EXCITON DIFFUSION LENGTH BASED ON GAUSSIAN APPROXIMATION

The average exciton diffusion length of our system can be calculated according to:

$$L_{\text{diff}} = \sqrt{\sigma_{\text{diff}}^2 - \sigma_{\text{no-diff}}^2},\tag{S1}$$

where $\sigma_{\text{diff}}^2$ is the exciton distribution variance in the presence of diffusion (measured on the closed-pack films) and $\sigma_{\text{no-diff}}^2$ is the exciton distribution variance without diffusion (measured on the sparse films). The PL intensity profile measured in a far-field microscopy system is given by the convolution of the single-emitter point spread function (PSF) with the excited-state population density (*i.e.* the exciton distribution). The PL intensity profiles and their underlying excited state population profiles are well approximated by Gaussian functions, so that the addition rule of variances upon convolution applies. The exciton diffusion length as defined above can be determined from the difference in the measured widths of the PL profiles since the effect of the PSF convolution cancels out. The steady-state intensity PL profile was measured on the close-packed and sparse samples repeating the measurement in multiple locations on the same sample and across several samples made from the same solution of PNCs. For each measurement the PL intensity profile was fitted with a Gaussian function and the profile width was extracted as the variance of the Gaussian fit. The two width distributions are shown in Supplemental Fig. S2 for the sparse and close-packed films; their mean values $\langle\sigma_{\text{close-packed}}^2\rangle$ and $\langle\sigma_{\text{sparse}}^2\rangle$ were used to calculate the average exciton diffusion length for our system according to

$$\langle L_{\text{diff}}\rangle = \sqrt{\langle\sigma_{\text{close-packed}}^2\rangle - \langle\sigma_{\text{sparse}}^2\rangle},\tag{S2}$$

and it was found to be 200 nm.

## S3. MODELING EXCITON PROCESSES

This supplement describes in detail the methods we used to model exciton hopping in nanoparticles. We approximated exciton transport within nanoparticle arrays as classical stochastic processes at mean-field (Section S4) and microscopic (Section S5) levels of detail, respectively. In Section S6 we compare simulation results with experiments.

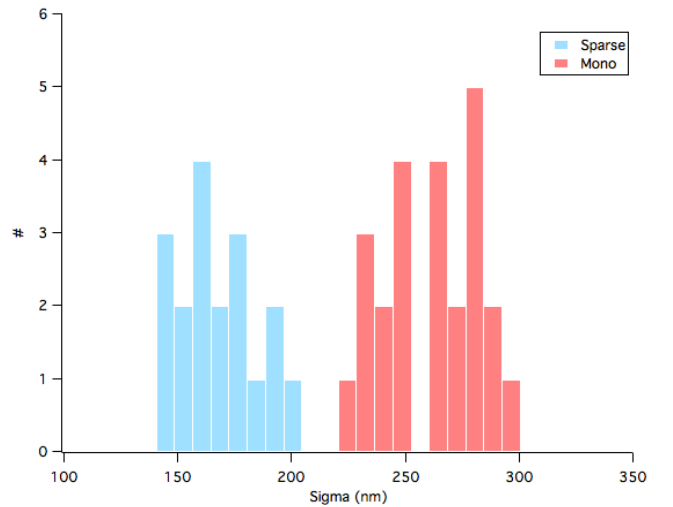

FIG. S2: Histogram of the PL profile sigma measured on a sparse monolayer of PNCs (light blue) and on a close-packed monolayer of PNCs (coral). The average values of the two distributions are (167±18) nm and (260±22) nm respectively.

Readers interested solely in a comparison between the experimental profiles described in the main text and simulated profiles should focus on Section S6. In this section the parameters used in simulations are chosen to match those of our experiments: we consider a square nanoparticle grid of lattice constant 10 nm; a laser source that is an Airy profile of full width half-maximum (FWHM) 240 nm; a laser source intensity low enough that no exciton-exciton interactions occur; a point-spread function for received light that is an Airy profile of FWHM 270 nm; excitons of lifetime ≈ 1 ns; and an exciton hopping rate such that their diffusion constant, on a pristine lattice, is ≈ 0.5 cm$^2$/s. By contrast, in Section S4 and Section S5, in which we describe in detail the simulation methods used, we use a variety of parameters, chosen for convenience or to make contact with results described in the literature. For instance, we sometimes approximate the laser source to be a Gaussian function, to illustrate differences with the Airy function, or we vary the exciton hopping rate or laser beam intensity in order to illustrate important trends that inform our understanding of the processes under study. We have done this for completeness and to provide the detail required to replicate the results described here.

To summarize, Sections S4 and S5 detail the methods used and make contact with results given in the literature; Section S6 contains the results specific to the present experiments.

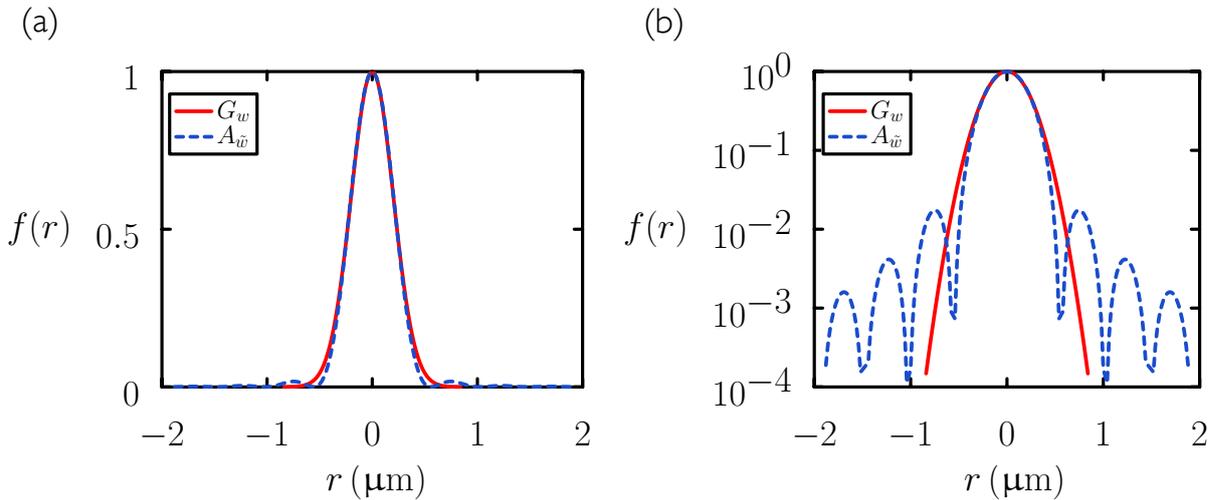

FIG. S3: The Gaussian function $G_w(r)$ and the Airy function $A_{\tilde{w}}(r)$ plotted on linear-linear (a) and linear-log (b) scales ($w = 0.2\ \mu$m). By setting $\tilde{w} \approx 0.728\,w$ we can arrange for the two functions to have equal widths at half their maximum value, and as seen in panel (a) the two functions indeed look similar on that scale. However, panel (b) shows how different the tails of the two functions are. These tails matter to the experiments described in this paper, because the point-spread function of our optics is an Airy function. This function is both the source of radiation 'felt' by the substrate-bound nanoparticles, and is the function convolved with the resulting exciton profile, via Eq. (S18), to form the observed profile. When computing the tails of steady-state profiles it is important not to approximate the point-spread function as a Gaussian: see Supplemental Fig. S7.

## S4. MEAN-FIELD EQUATIONS

### A. Effective description of exciton behavior

The creation, hopping, and recombination of laser-induced excitons in nanoparticles results from the quantum mechanics of light-matter interactions. In this section we approximate these processes by considering the statistics of classical bosonic particles. Consider the diffusion, creation, self-destruction, and pair annihilation of classical bosonic particles on a lattice (see e.g. Ref. [2]). Writing down a master equation for this set of processes, taking the continuum steady-state limit, and ignoring fluctuating noise terms, we get

$$D\nabla^2 c(\mathbf{x}) - \mu c(\mathbf{x}) - \rho c(\mathbf{x})^2 = -\Phi(\mathbf{x}). \quad (S3)$$

Here $c(\mathbf{x})$ is the concentration (number per unit area) of particles (excitons) at spatial location $\mathbf{x} = (x, y)$ on a two-dimensional substrate. We shall regard (S3) as an effective steady-state description of the optical signal produced by laser-induced nanoparticle excitons; to do so we make the additional assumption that photons are emitted by isotropic one-body exciton decay, i.e. that the optical signal at position $\mathbf{x}$ is proportional to $c(\mathbf{x})$. This description is approximate in several respects, as we shall describe, but several features of its solution, and in particular the approximate shape of the exciton profile produced, provide insight into the workings of our experiments.

The term in (S3) that couples to $D$ describes the diffusion of excitons. This is an approximation: exciton hopping is generally sub-diffusive on small lengthscales and timescales [3]. The term in $\mu$ describes the self-destruction of excitons. The term in $\rho$ describes pair annihilation of (bosonic) particles. A similar term (plus higher-order nonlinearities) would also be present in an effective description if particles are instead fermionic (i.e. if only one exciton per nanoparticle is permitted). We address this case in Section S5.

The term $\Phi(\mathbf{x})$ describes the intensity of the laser beam at position $\mathbf{x} = (x, y)$ on the two-dimensional substrate. We shall consider profiles with radial symmetry in the plane, and we will write $\Phi(\mathbf{x}) = \Phi_0 f(r/w)$. Here $\Phi_0$ is proportional to the laser power output; $f(r/w)$ is a function containing the laser beam width parameter $w$; and $r \equiv \sqrt{x^2 + y^2}$ is the radial coordinate. We shall consider cases in which $f$ is an Airy function or its Gaussian approximation. The Airy function is

$$A_w(r) \equiv \left(\frac{2J_1(r/w)}{r/w}\right)^2, \quad (S4)$$

where $J_1$ is the first-order Bessel function of the first kind. The parameter $w$ quantifies the width of the laser beam. The Gaussian function is

$$G_w(r) \equiv \exp\left(-\frac{r^2}{2w^2}\right). \quad (S5)$$

The functions $G_w(r)$ and $A_{\tilde{w}}(r)$ have equal widths at half their maximum value when $\tilde{w} \approx 0.728\,w$, but differ substantially in their tails; see Supplemental Fig. S3. In this supplement we often use the Gaussian profile for the purposes of illustration. When comparing with experimental data, however, we use the Airy function.

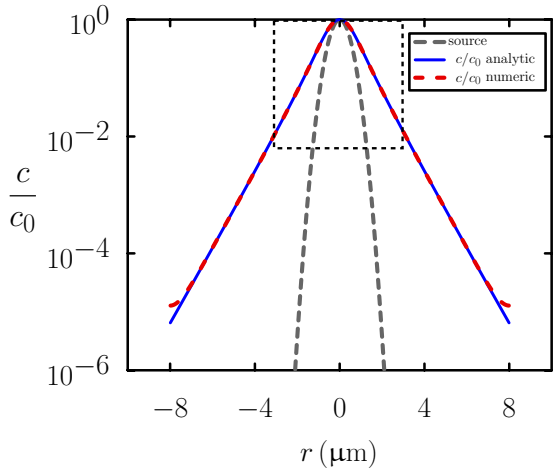

FIG. S4: Numerical (dashed red) [Eq. (S11)] and semi-analytic (blue) [Eq. (S15)] solutions to the linearized version of Eq. (S3) agree, providing a check on our numerical procedure. The parameter combinations are $\mathcal{P} = 0$ and $\mathcal{R}^2 \equiv \mu w^2 / D = 0.32$; see Section S4 B. Exciton concentration curves have been normalized by their value at the origin, $c(0) \equiv c_0$. The (normalized) model source is shown grey; for simplicity it is Gaussian with width parameter $w = 0.4$ µm. The dotted box shows the region relevant to typical experimental measurements, in which the measured intensity spans two or three orders of magnitude.

## B. Scaling analysis

A scaling analysis shows that (S3) is governed by two parameter combinations. Eq. (S3) can be written

$$\nabla^2 c - \left(\frac{\rho}{D}\right) c^2 - \left(\frac{\mu}{D}\right) c = -\frac{\Phi_0}{D} f\left(\frac{r}{w}\right). \quad (S6)$$

If we choose to measure lengths in units of the beam width $w$, and introduce coordinates $(\hat{x}, \hat{y}) \equiv w^{-1}(x, y)$, then (S6) becomes

$$\hat{\nabla}^2 c - \left(\frac{\rho w^2}{D}\right) c^2 - \left(\frac{\mu w^2}{D}\right) c = -\frac{\Phi_0 w^2}{D} f(\hat{r}). \quad (S7)$$

Introducing a rescaled concentration field *via* $c \equiv \left(\Phi_0 w^2 / D\right) \hat{c}$ brings (S7) to the form

$$\hat{\nabla}^2 \hat{c} - \mathcal{P} \hat{c}^2 - \mathcal{R}^2 \hat{c} = -f(\hat{r}). \quad (S8)$$

Here we have introduced the parameter combinations

$$\mathcal{P} \equiv \frac{\rho \Phi_0 w^4}{D^2} \quad (S9)$$

and

$$\mathcal{R}^2 \equiv \frac{\mu w^2}{D}. \quad (S10)$$

The combination $\mathcal{P} \equiv \rho \Phi_0 w^4 / D^2$ is the nonlinearity or power parameter. When $\mathcal{P}$ is large the beam is powerful in the sense that the term nonlinear in $c$ is important. When $\mathcal{P}$ is small we are in the linear regime, where the term in $\rho$ in Eq. (S3) may be ignored. The experiments reported in this paper are performed in the linear regime.

The parameter $\mathcal{R} \equiv \sqrt{\mu w^2 / D}$ is a ratio of lengthscales. $\ell_{\text{beam}} \equiv w$ is the lengthscale associated with the beam profile. $\ell_{\text{hop}} \equiv \sqrt{D/\mu}$ is the lengthscale on which an exciton that lives for characteristic time $\mu^{-1}$ will hop before it dies. Thus when $\mathcal{R} = \ell_{\text{beam}}/\ell_{\text{hop}}$ is large, the beam diameter is much greater than the distance over which a typical exciton will diffuse. When $\mathcal{R}$ is small, the beam diameter is much less than the exciton hopping distance. For the experiments reported in the main text the lengths $\ell_{\text{beam}}$ and $\ell_{\text{hop}}$ are comparable.

## C. Numerical solution of Eq. (S3)

In general, Eq. (S3) must be solved numerically. To do so we simulated numerically the time-dependent version of the equation on a 2D periodic grid, using a forward-different method and a five-point Laplacian stencil:

$$
\begin{aligned}
c_{x,y}(t + \Delta t) \;=\; & c_{x,y}(t) + \Delta_t \left[ -\rho c_{x,y}(t)^2 - \mu c_{x,y}(t) + \Phi_0 f(r\Delta_x^{-2}/w) \right] \\
& + D\Delta_t \Delta_x^2 \left[ c_{x+1,y}(t) + c_{x-1,y}(t) + c_{x,y+1}(t) + c_{x,y-1}(t) - 4c_{x,y}(t) \right].
\end{aligned}
\quad (S11)
$$

Here $c_{x,y}(t)$ is the exciton concentration at grid point $(x, y)$ at time $t$. We measured spatial distances in microns, and took $\Delta_x = 50$ (*i.e.* we have 50 lattice spacings to the micron). We set the timestep $\Delta t = 10^{-5}$, which was small enough to maintain numerical stability. We usually began simulations with a spatial profile $c(r)$ equal to that of the source term $\Phi_0 f(r\Delta_x^{-2}/w)$, and sim-ulated until the steady state was reached. We confirmed the accuracy of our numerics by comparing the steady-state solution of Eq. (S11) to a semi-analytic solution of Eq. (S3) in a certain limit (Section S4 D): see Supplemental Fig. S4.

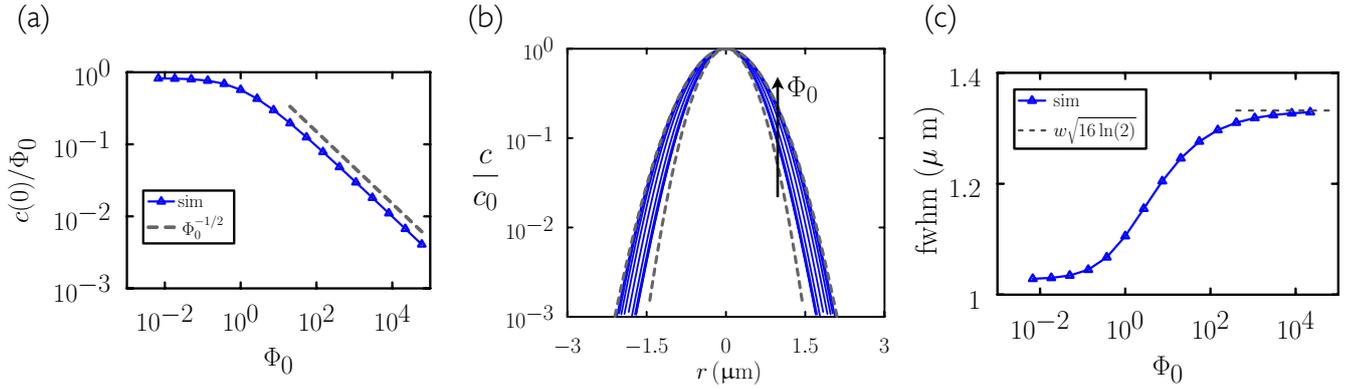

FIG. S5: Numerical solutions to Eq. (S3) for a range of source intensities $\Phi_0$ show that the scale and shape of profiles change as we move from the linear to the nonlinear regime. For the purposes of illustration we consider a Gaussian source with width parameter $w = 0.4$ μm. (a) The maximum exciton concentration changes dependence upon $\Phi_0$ upon moving from the linear to the nonlinear regime. (b) The profile widths tend towards the square root of the souce profile (the outer dotted line) in the large-$\Phi_0$ limit. (c) The full widths at half-maximum value (FWHM) of the profiles change accordingly. Equation parameters: the lengthscale parameter combination (S10) is $\mathcal{R} = \mu w^2/D = 8$, indicating that the source width is greater than the exciton diffusion length (and so low-power profiles are only slightly broader than the source). The power parameter (S9) is $\mathcal{P} = 64\Phi_0$, which must be much smaller than unity to be in the linear regime.

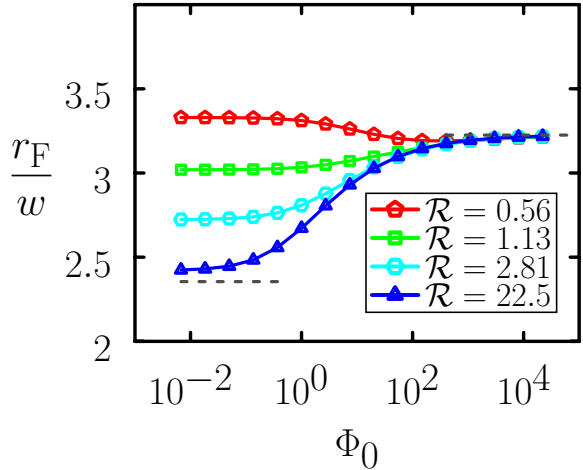

FIG. S6: The full width at half maximum $r_F$ of a series of normalized exciton profiles, plotted relative to the source width $w$ (= 0.15 μm), for a range of source intensities $\Phi_0$. Distinct curves correspond to distinct choices of exciton diffusion constant $D$; the resulting dimensionless parameters $\mathcal{R} \equiv w^2\mu/D$ are shown. The dotted lines left and right indicate the width of the source and $\sqrt{2}$ times that value, respectively. The smaller is $\mathcal{R}$ the broader is the exciton profile in the linear (small-$\Phi_0$) regime. At large $\Phi_0$, in the strongly nonlinear regime, all profile widths tend to a value $\sqrt{2}$ times that of the source. Whether nonlinear broadening or narrowing occurs depends therefore on the value of $\mathcal{R}$.

### D. Semi-analytic solution of the linearized version of Eq. (S3)

As a benchmark for our numerics (Section S4 C) and to gain insight into the properties of Eq. (S3), it is instruc-tive to solve the equation in the absence of the term in $\rho$, *i.e.* in the linear limit. The linear limit is appropriate when the source intensity $\Phi_0$ and the resulting maximum exciton concentration is small.

In 2D the solution of Equation (S3) can be obtained by the method of Green's functions, and is

$$c(x,y) = \frac{\Phi_0}{2\pi D}\int_{-\infty}^{\infty}dy'\int_{-\infty}^{\infty}dx'\Phi\left(\sqrt{(x')^2+(y')^2}\right)$$
$$\times\ K_0\left(\lambda\sqrt{(x-x')^2+(y-y')^2}\right). \quad (S12)$$

Here $\lambda \equiv \sqrt{\mu/D}$ is the reciprocal of the characteristic lengthscale for exciton diffusion, and $K_0$ is the zeroth order modified Bessel function of the second kind. Writing $u \equiv x - x'$ and $v \equiv y - y'$, and passing to plane polar coordinates *via* the transformations $(u,v) = \xi(\cos\theta', \sin\theta')$ and $(x,y) = r(\cos\theta, \sin\theta)$ gives

$$c(r,\theta) = \frac{\Phi_0}{D}\int_0^{\infty}\xi d\xi K_0(\lambda\xi)$$
$$\times\ \int_0^{2\pi}\frac{d\theta'}{2\pi}\Phi\left(\sqrt{r^2+\xi^2+2r\xi\cos(\theta-\theta')}\right). \quad (S13)$$

The source $\Phi$ in our experiments is an Airy function, but for the purposes of checking our numerics we replace it by a Gaussian function. In this case (S13) can be reduced to a single integral. Setting $\Phi(r) = \Phi_0 e^{-(x^2+y^2)/(2w^2)}$ we have

$$c(r,\theta) = \frac{\Phi_0}{D}e^{-r^2/(2w^2)}\int_0^{\infty}d\xi\,\xi e^{-\xi^2/(2w^2)}K_0(\lambda\xi)$$
$$\times\ \int_0^{2\pi}\frac{d\theta'}{2\pi}e^{r\xi w^{-2}\cos(\theta-\theta')}. \quad (S14)$$

The inner integral can be carried out using the formula $\int_0^{2\pi}d\theta\exp(\alpha\cos\theta+\beta\sin\theta) = 2\pi I_0(\sqrt{\alpha^2+\beta^2})$, where $I_0$ is the zeroth-order modified Bessel function of the first kind. This result allows us to write (S14) in the manifestly $\theta$-independent form

$$
\begin{aligned}
c(r) \;=\;& \frac{\Phi_0}{D} \mathrm{e}^{-r^2/(2w^2)} \\
& \times \int_0^\infty \mathrm{d}\xi\, \xi \mathrm{e}^{-\xi^2/(2w^2)} K_0\left(\lambda \xi\right) I_0\left(\frac{r\xi}{w^2}\right),
\end{aligned}
\tag{S15}
$$

which we can evaluate numerically. In Supplemental Fig. S4 we show that, in the relevant parameter regime, the steady-state limit of the numerical procedure (S11) agrees with the semi-analytic solution (S15), except near the edge of the simulation box (where artifacts associated with periodic boundaries are apparent). The dotted square shows the scale on which we typically compare our simulations with experimental data, which is well away from such artifacts.

Note that rescaling space and the concentration field in the manner described in Section S4 B confirms that (S15) depends only upon the single parameter combination $\mathcal{R} \equiv \sqrt{\mu w^2/D}$:

$$
\hat{c}(\hat{r}) = \mathrm{e}^{-\hat{r}^2/2} \int_0^\infty \mathrm{d}\hat{\xi}\, \hat{\xi} \mathrm{e}^{-\hat{\xi}^2/2} K_0\left(\hat{\xi}\mathcal{R}\right) I_0\left(\hat{r}\hat{\xi}\right). \tag{S16}
$$

### E. Interpolation between linear and nonlinear regimes

When the nonlinear term is present in Eq. (S3), the shape of the exciton profile changes with source power $\Phi_0$. Consider a Gaussian source of width $w$, $\Phi(r) = \Phi_0 G_w(r)$ [see Eq. (S5)]. When $\Phi_0$ is large, such that the parameter combination $\mathcal{P} \equiv \rho\Phi_0 w^4/D^2 \gg 1$, Eq. (S3) can be approximated near its core as

$$
\rho c^2 = \Phi_0 G_w(r), \tag{S17}
$$

from which we get $c(0) \propto \Phi_0^{1/2}$. Thus, as shown in Supplemental Fig. S5(a), plotting $c(0)/\Phi_0$ for a series of numerical experiments carried out at different source intensities $\Phi_0$ indicates the onset of the nonlinear regime. The value of the gradient of the function in the high-power regime depends on the types of nonlinearities present (*e.g.* it differs for bosonic and fermionic excitations), but the qualitative change can be used to determine the extent of the linear regime.

As we move from the linear to the nonlinear regime, the width of the exciton profile changes. In the linear regime the width of the exciton profile is determined (for the model diffusion equation) by the source width $w$ and the decay length $\sqrt{D/\mu}$. In Supplemental Fig. S5(b) we show a series of normalized exciton profiles (blue) that result from (S3), for the range of choices of source power $\Phi_0$ shown in panel (a). In (b), the inner profile corresponds to the case of lowest power, and is slightly larger than the source (the inner dotted gray line) by virtue of the diffusive broadening seen in Supplemental Fig. S4. As source intensity increases, the profiles broaden. From Eq. (S17) we see that in the limit of large intensity, the spatial profile has the shape $c(r) \propto G_w(r)^{1/2} = G_{w\sqrt{2}}(r)$, which is a Gaussian with a width $\sqrt{2}$ times that of the source. As shown in Supplemental Fig. S5(b), the outer blue profiles indeed tend to this shape (shown by the outer dotted gray line). Plotting profiles' full width at half maximum value in Supplemental Fig. S5(c), we see that they tend to the expected width in the limit of large $\Phi_0$.

Nonlinear *narrowing* can also be seen, when the decay length $\sqrt{D/\mu}$ is large compared with the source width $w$. In this case the dimensionless parameter $\mathcal{R} < 1$, and the profile width in the linear regime can be broader than the profile width in the strongly nonlinear regime, which tends to a value $\sqrt{2}$ times that of the source; see Supplemental Fig. S6. Note that these width comparisons refer to shapes of *normalized* profiles, those scaled by their values $c(0)$ at the origin: profiles generated at large values of $\Phi_0$ are generally broader than those generated at small $\Phi_0$, in the sense that greater exciton density is generated away from the origin.

### F. Observed profiles are a convolution of the exciton profiles and the optics' point-spread function

Intensity profiles $I(r)$ observed in experiment are not the exciton profiles $c(r)$ themselves, but are instead the convolution of the exciton profile and the point-spread function $S(r)$ of the optics [3]:

$$
\begin{aligned}
I(r) \;=\;& (c \star S)(r) \\
=\;& \int \mathrm{d}x'\mathrm{d}y'\, c(x',y') S(x-x', y-y').
\end{aligned}
\tag{S18}
$$

The optics plays a dual role in our experiments: it gives rise to an Airy-function source profile $\Phi(r)$ of FWHM 240 nm on the substrate, and it gives rise to the point-spread function $S(r)$ for received light, an Airy function of FWHM 270 nm, that appears in (S18). Airy functions are sometimes approximated as Gaussian functions, because the cores of the two profile types have similar shapes [Supplemental Fig. S3(a)]. However, the tails of the two functions differ markedly [Supplemental Fig. S3(b)]. This difference is significant when computing steady-state profiles, as shown in Supplemental Fig. S7, particularly as regards inflation of the tail of the profile. We carried out the convolution (S18) numerically.

The trends described previously, such as the broadening of profiles at large source power, can be seen in $I(r)$ much as in $c(r)$, with quantitative differences: see *e.g.* Supplemental Fig. S8.

## S5. MICROSCOPIC SIMULATIONS

To complement the approach of Section S4 we simulated space-dependent exciton dynamics using discrete-time and continuous-time Monte Carlo algorithms.

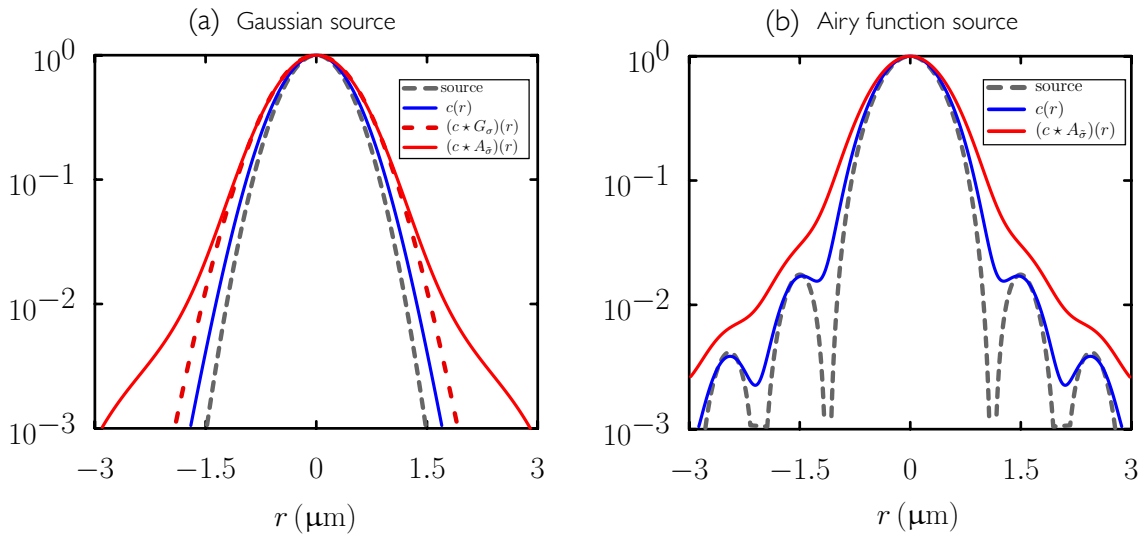

FIG. S7: Gaussian and Airy functions used for the source profile $\Phi(r)$ and the convolution (S18). (a) Numerical solution to Eq. (S3) for a Gaussian source $\Phi(r) \propto G_\sigma(r)$ of width $\sigma = 0.4$ μm (gray dotted) gives an exciton profile $c(r)$ (blue). Subsequent convolution with a Gaussian (red dotted) or an Airy function (red solid) produce distinct curves. (b) Numerical solution to Eq. (S3) for an Airy function source $\Phi(r) \propto A_{\tilde\sigma}(r)$, with $\tilde\sigma = 0.7\sigma$ (gray dotted), gives an exciton profile $c(r)$ (blue) whose tails are markedly different to the tails of $c(r)$ with a Gaussian source. Subsequent convolution with an Airy function gives the solid red line.

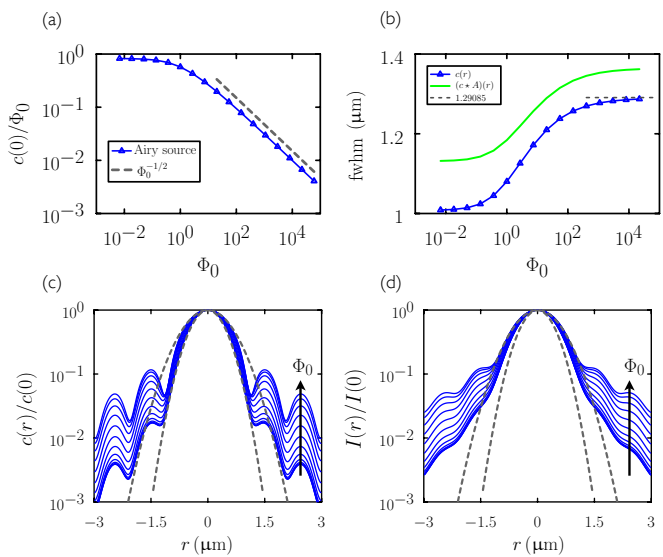

FIG. S8: Observed profiles $I(r)$ vary in a similar fashion to exciton profiles $c(r)$, with quantitative differences. (a) Similar to Supplemental Fig. S5(a), but with an Airy-function source of width 0.24 μm. (b) The full width at half maximum (FWHM) of exciton- and observed profiles behaves similarly, but are numerically different. Profiles $c(r)$ and $I(r)$ are shown in (c) and (d), respectively, overlaid on Gaussian reference curves.

## A. Fermionic statistics

We consider a two-dimensional substrate of nanoparticles whose positions are fixed. In Section S6 we take the nanoparticles to sit in a square array, similar to the experiments reported in the main text; in this section we also consider trangular arrays and disordered arrangements. We model excitons as classical particles, able to undergo various processes. Each nanoparticle can be occupied by an exciton A or be vacant $\emptyset$, *i.e.* we assume fermionic exciton statistics. In this case the normalized exciton profiles broaden at high power even in the absence of exciton hopping. The broadening is different in detail to that of the bosonic statistics considered in Section S4. Consider exciton creation with rate $\Phi(r)$,

$$\emptyset \xrightarrow{\Phi(r)} A, \tag{S19}$$

where $\Phi(r) = \Phi_0 f(r)$ is the laser source as in Section S4, and exciton self-destruction with rate $\mu$,

$$A \xrightarrow{\mu} \emptyset. \tag{S20}$$

The stochastic process defined by (S19) and (S20) is a two-state dynamics with steady-state solution

$$c(r) = \frac{\Phi(r)}{\Phi(r) + \mu}, \tag{S21}$$

where $c(r)$ is the density of excitons (A-particles) at position $r$. We assume that the process of destruction produces a photon, and so the time-averaged exciton density

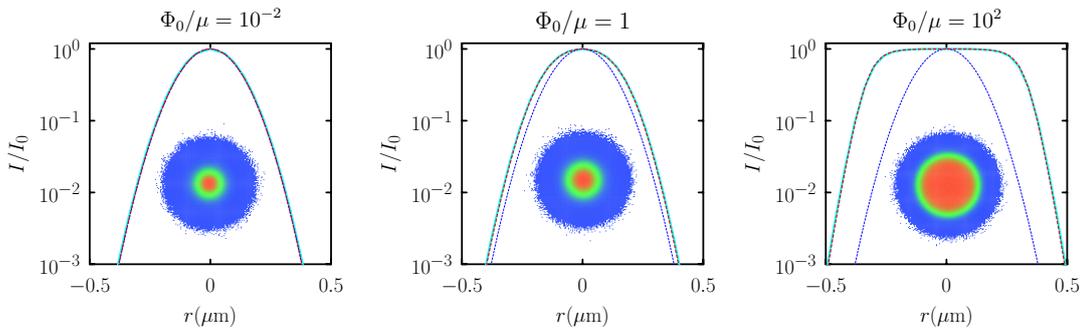

FIG. S9: Numerically computed steady-state radial exciton profiles $c(r)$ (cyan) and the exact solution (S21) (red dashed) for the processes (S19) and (S20), together with the model Gaussian source profile (blue dotted). Shown inset are the two-dimensional images from which the radial profiles are computed.

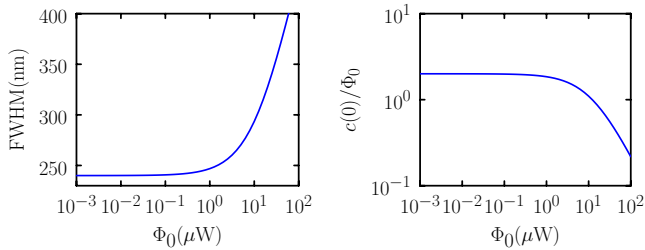

FIG. S10: FWHM (left) and normalized intensity (right) of exciton profiles $c(r)$ produced by the stochastic processes (S19) and (S20), for varying beam power $\Phi_0$. We take $\mu = 0.5\,\mathrm{ns}^{-1}$. Note that the FWHM continues to broaden with source power, unlike the case of bosonic exciton statistics.

is proportional to the steady-state photoluminescence intensity.

We simulated these processes using continuous-time Monte Carlo [4] and a square-lattice nanoparticle array. Comparison with (S21) provides a simple benchmark against which to check the calculation of radially-averaged profiles. As shown in Supplemental Fig. S9, the time- and radially-averaged profile $c(r)$ is proportional to the source profile $\Phi(r)$ at low power, and broadens as beam power is increased. For the purposes of illustration we take the source $\Phi(r) = \Phi_0 G_w(r)$ to be Gaussian with full width at half maximum intensity (FWHM) 240 nm, i.e. $w = (120/\sqrt{2\ln 2})\,\mathrm{nm}$.

For fermionic statistics the width of the profile grows logarithmically with power at high power: we can solve the equation $c(r_0) = c(0)/2$ to yield the FWHM, $F \equiv 2r_0$:

$$F(\Phi_0) = 2w\sqrt{2\ln\left(\frac{\Phi_0}{\mu} + 2\right)}. \qquad (S22)$$

We can also calculate the total integrated intensity

$$I_{\mathrm{tot}} = \int \mathrm{d}\theta\, r\mathrm{d}r\, c(r) = 2\pi w^2 \ln\left(\frac{\mu + \Phi_0}{\mu}\right), \qquad (S23)$$

for Gaussian $\Phi(r)$. We have $I_{\mathrm{tot}} \approx 2\pi w^2 \Phi_0/\mu$ for small $\Phi_0/\mu$. Equations (S22) and (S23) are plotted in Supplemental Fig. S10. We take the rate of self-destruction to be $\mu = 0.5\,\mathrm{ns}^{-1}$ and the beam power parameter to be $\Phi_0 = (P/24.4)\,\mathrm{ns}^{-1}$, where $P$ is measured in microwatts ($\mu$W), which we estimate to be characteristic of our experiments. These behaviors are useful diagnostics of the onset of nonlinear behavior, and allow us to verify that experiments reported in the main text are done in the linear regime.

## B. Exciton hopping is subdiffusive in the presence of energetic disorder

In the experiments reported in the main text we believe that subdiffusive motion of excitons arises from vacancies in the nanoparticle array. In this section we recall some features of exciton subdiffusion brought about by another mechanism, energetic disorder, that has been quantified in other studies [3]. Hopping on a rough energy landscape leads in general to subdiffusive behavior at short times and diffusive behavior at long times [5, 6].

To make contact with these results we carried out Monte Carlo simulations of an exciton moving between nanoparticles on a two-dimensional substrate; see Supplemental Fig. S11. Simulation boxes had periodic boundaries in both dimensions. We considered spatially ordered substrates, in which nanoparticles were arranged as a close-packed lattice with inter-particle separation $a = 8$ nm – see Supplemental Fig. S11(a) – and spatially disordered substrates, such as that shown in Supplemental Fig. S11(b). These we generated by performing short constant-volume Monte Carlo simulations of the nanoparticles themselves, assuming they were hard discs with radii drawn from a truncated Gaussian distribution peaked about 8 nm. We found (shown below) that at constant particle density the averaged exciton transport properties were not strongly affected by the presence of spatial disorder.

Once the substrate was generated, we performed ex-citon hopping simulations using a discrete-time Monte Carlo algorithm. We selected at random a nanoparticle, and created an exciton on that nanoparticle. We then selected at random any neighbor (up to a cutoff distance) of that nanoparticle, and proposed to move the exciton to that nanoparticle. Following Ref. [7] we accepted this proposal with a probability designed to ensure that the exciton jump from $i$ to $j$ happens with rate

$$R(i \to j) = \frac{1}{\tau_0} \left( \frac{R_0}{R_{ij}} \right)^6 \min \left( 1, \mathrm{e}^{-\beta(E_j - E_i)} \right). \quad (S24)$$

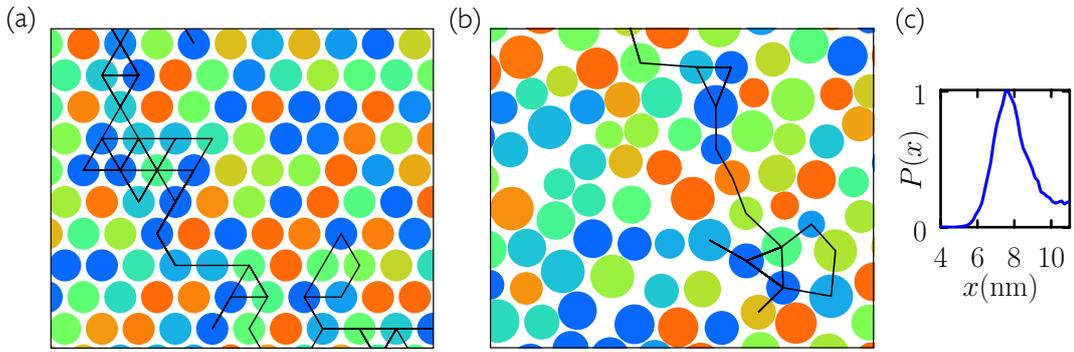

FIG. S11: Examples of spatially ordered (a) and disordered (b) substrates used for illustrative exciton-hopping simulations. The black traces show trajectories taken by two simulated excitons. Nanoparticle colors indicate their energies; red shades and blue shades are high and low in energy, respectively. Panel (c) shows the distribution of inter-nanoparticle distances seen in panel (b).

Here $i$ and $j$ are the nanoparticle identities; $\tau_0$ is the mean exciton lifetime; $R_0$ is the Förster radius; $R_{ij}$ is the distance between nanoparticles $i$ and $j$; $E_i$ and $E_j$ are the bandgaps of nanoparticles $i$ and $j$; and $\beta \equiv (k_B T)^{-1}$. For spatially disordered substrates the combination $R_0/R_{ij}$ can be greater than or less than unity, and so it is convenient to write (S24) as

$$R(i \to j) = \frac{1}{\tau} \left( \frac{R_{\min}}{R_{ij}} \right)^6 \min \left( 1, \mathrm{e}^{-\beta(E_j - E_i)} \right), \quad (S25)$$

with $R_{\min} \equiv \min_{ij} R_{ij}$ and $\tau \equiv \tau_0 (R_{\min}/R_0)^6$. With time measured in units of $\tau$ we accepted the move from $i$ to $j$ with probability

$$p_{\mathrm{acc}}(i \to j) = \left( \frac{R_{\min}}{R_{ij}} \right)^6 \min \left( 1, \mathrm{e}^{-\beta(E_j - E_i)} \right), \quad (S26)$$

which is $\leq 1$. Otherwise, the proposed exciton move was rejected. We considered a Gaussian distribution of nanoparticle energy levels $E_i$ with variance $\epsilon^2$, $P(E_i) \propto \exp\left(-E_i^2/(2\epsilon^2)\right)$.

The exciton diffusion parameter is

$$D(t) = \frac{\langle [\Delta x(t)]^2 \rangle + \langle [\Delta y(t)]^2 \rangle}{4N_{\mathrm{steps}}} \cdot \frac{a^2}{\tau_0} \left( \frac{R_{\min}}{R_0} \right)^6, \quad (S27)$$

where $\Delta x(t)$ and $\Delta y(t)$ are the time-dependent distances (in units of $a$) traveled in each dimension by excitons

(corrected for periodic boundaries); averages $\langle \cdot \rangle$ are taken over initial conditions, waiting times and (where appropriate) realizations of energetic and spatial disorder; and $N_{\mathrm{steps}}$ is the number of Monte Carlo steps taken. Simple considerations indicate roughly the exciton diffusion constant expected. Take the nanoparticle radius to be $a \sim 10$ nm. Assume the characteristic rate for an exciton to hop from nanoparticle to nanoparticle is $\tau_0^{-1}(R_0/a)^6$, where $\tau_0 \sim 10$ ns, and assume that the Förster radius $R_0$ is of order 10 nm [3]. Then the long-time exciton diffusion constant is roughly $D = \frac{1}{2\tau_0} \left( \frac{R_0}{a} \right)^6 a^2 \approx 10^{-4}$ cm$^2$/s. This scale of this result is consistent with the exciton diffusion constant of $3 \times 10^{-4}$ cm$^2$/s reported in Ref. [3]; the precise numerical value of this result is sensitive to the ratio $R_0/a$ to the sixth power, and upon insertion of different values (e.g. $R_0 = 12.5$ nm and $a = 8$ nm) we obtain the numbers shown in Supplemental Fig. S12(a).

In that figure we show $D$ from Eq. (S27), as a function of time, for four different values of energetic disorder (on a spatially uniform lattice). A constant value indicates diffusive motion, which is reached at times that increase as the roughness $\epsilon/(k_B T)$ of the energy landscape increases. For e.g. nanoparticles for which $\epsilon \approx k_B T$, we estimate the diffusive approximation made in Eq. (S3) to be valid only on timescales exceeding about 100 ns [in general the $D$ appearing in (S3) could be thought of as a spatial and temporal average over the microscopic behavior shown in Supplemental Fig. S12(a)].

In Supplemental Fig. S12(b) we show the long-time diffusion constant obtained for particular values of substrate energetic disorder, normalized by the value for no energetic disorder. For a spatially ordered lattice (green line), the fall-off of $D$ with $\epsilon/(k_B T)$ is less rapid than for discrete Gaussian disorder on a 1D lattice (blue dotted line), $D(\epsilon)/D(0) = \exp(-\beta^2 \epsilon^2)(1 + \mathrm{erf}(\beta \epsilon/2))^{-1}$ [6]. This makes physical sense, because energetic 'traps' caused by the proximity of nanoparticles with unusually high and low energies are geometrically harder to avoid in 1D than in 2D. The simulation result also shows a more rapid fall-off with $\epsilon/(k_B T)$ does $D$ for a continuous Gaussian surface (solid blue line), $D(\epsilon)/D(0) = \exp(-\beta^2\epsilon^2/2)$ [6]. This hierarchy also makes physical sense: a continuous surface is less likely to give rise to particularly abrupt energy changes (traps) than are discrete energy levels drawn from a Gaussian distribution.

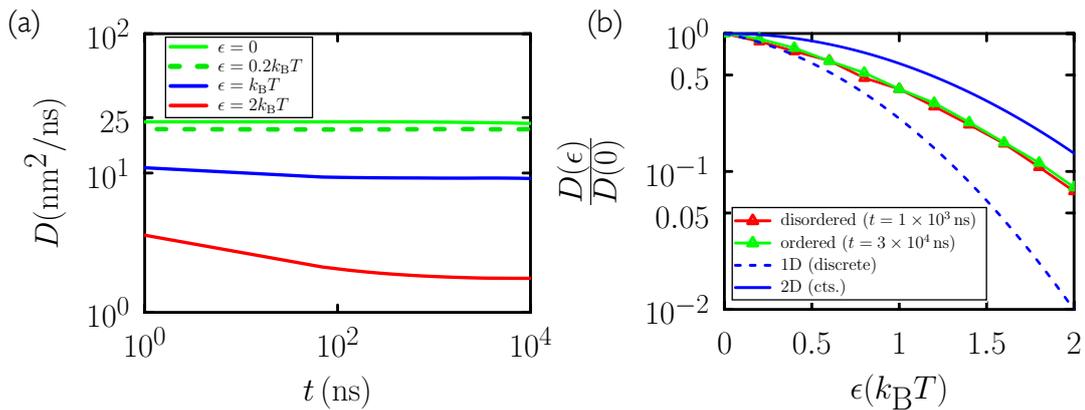

FIG. S12: (a) The diffusion parameter (S27) as a function of time, for four different values of energetic disorder (on a triangular lattice). A constant value indicates diffusive motion, which is reached at times that increase as the roughness $\epsilon/(k_B T)$ of the energy landscape increases. (b) Long-time diffusion constant obtained for particular values of substrate energetic disorder, normalized by the value for no energetic disorder. For a triangular lattice (green line) the fall-off of $D$ with $\epsilon/(k_B T)$ is less rapid than for discrete Gaussian disorder on a 1D lattice (dotted blue line), but more rapid than for a continuous Gaussian surface in 2D (solid blue line) [6]. The presence (red line) or absence (green line) of nanoparticle spatial disorder (at constant area) has little effect on the fall-off of $D$ with energetic roughness.

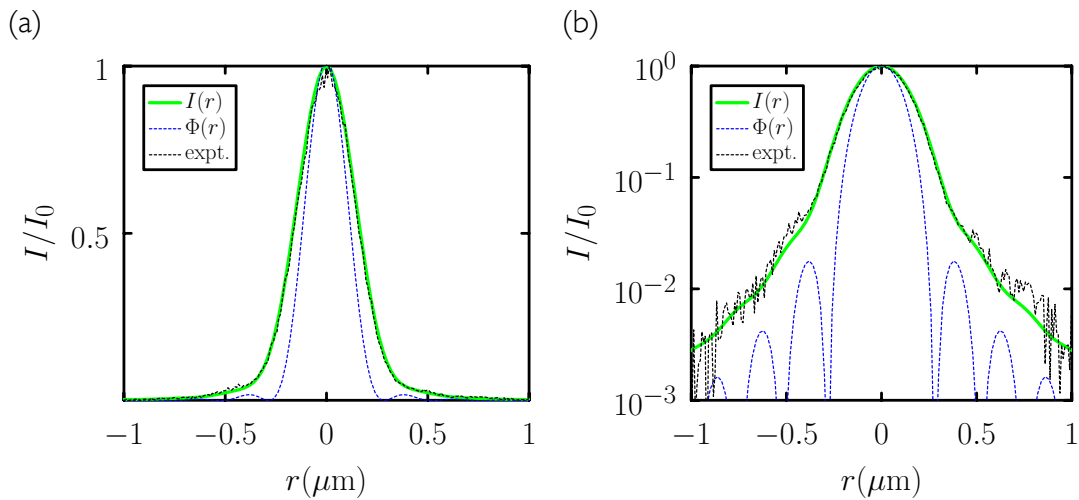

FIG. S13: The observed photoluminescence profile (black dashed line) on a sparse nanoparticle substrate is a convolution of our laser source $\Phi(r)$ (blue line), which is an Airy function with FWHM 240 nm, with the point-spread function of the optics for received light, which is an Airy function of FWHM 270 nm. The green line is the numerical convolution $I(r)$ of these two Airy functions, Eq. (S18). This convolution matches the observed profile, providing a baseline from which we can assess the effect of exciton hopping [see Supplemental Fig. S14]. Panels (a) and (b) show linear-linear and linear-log plots, respectively.

We found that the presence or absence of spatial disorder of nanoparticles (at constant nanoparticle areal coverage) has little effect upon $D(\epsilon)$ [compare red and green lines in Supplemental Fig. S12(b)]; the same is not true of spatial disorder at varying nanoparticle coverage, as described in Section S6.

## S6. COMPARISON WITH EXPERIMENTAL DATA

In this section the parameters used in simulations are chosen to match those of our experiments. We consider a square nanoparticle grid of lattice constant 10 nm (or a continuum approximation thereof); a laser source that is an Airy profile of full width half-maximum (FWHM) 240 nm; a laser source intensity low enough that no exciton-exciton interactions occur (we are in the linear regime); a point-spread function for received light that is an Airy profile of FWHM 270 nm; and excitons of lifetime $\approx 1$ ns.

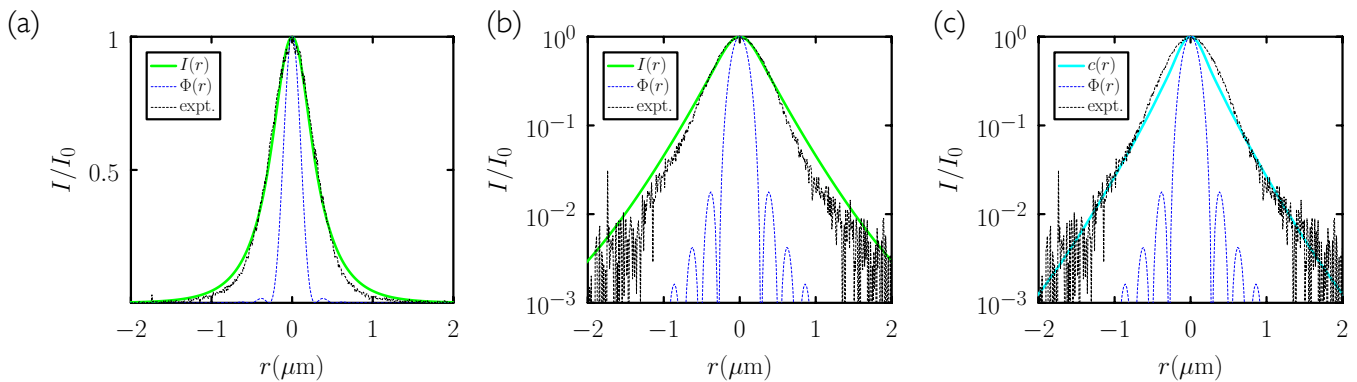

FIG. S14: The observed photoluminescence profile (black dashed line) on a dense nanoparticle substrate is broader than that shown in Supplemental Fig. S13, on account of exciton hopping. The source $\Phi(r)$ here is as in Supplemental Fig. S13. The green line $I(r)$ results from the solution $c(r)$ of Eq. (S3), for $D \approx 1\,(\mathrm{cm})^2/\mathrm{s}$, convolved with the point-spread function $S(r)$, per (S18). The scale of the broadening is consistent with that seen in experiment, supporting our estimate of the basic rate of exciton hopping. However, profile shapes are not identical, indicating that exciton hopping in experiment is not perfectly described by the diffusion equation. Panels (a) and (b) show linear-linear and linear-log plots, respectively. Panel (c) shows the exciton profile $c(r)$ (cyan line) that results from Eq. (S3) (this profile is not expected to match the experimental profile, because the latter involves convolution with the optics' point-spread function).

In Supplemental Fig. S13 we show the experimentally measured photoluminescence profile on a sparse nanoparticle substrate (black). Here we expect the rate of exciton hopping to be effectively zero, and so we can use this case as a baseline to isolate the effect of our optics. Also shown in the figure are an Airy function of FWHM 240 nm (blue), which is the profile of the laser source on the substrate, and (in green) the convolution of this function with an Airy function of FWHM 270 nm. The latter is the point-spread function of the optics at the received wavelength. The convolution matches the observed profile, even into the tails, indicating that our optics functions as expected. Knowing this baseline is important, because it allows us to attribute the broadening of the profile seen on the dense nanoparticle substrate to exciton hopping. We show the experimentally measured photoluminescence profile on a dense nanoparticle substrate (black) in Supplemental Fig. S14. We also show the profile $I(r)$ expected for diffusive excitonic motion (green). This profile results from the exciton profile $c(r)$, calculated from Eq. (S3) in the low-power regime $\rho = 0$ for $D/\mu = 0.095\,(\mu m)^2$, convolved, per (S18), with the point-spread function of the optics. In that equation, $S(r)$ is an Airy function of FWHM 270 nm. The exciton profile itself is shown in cyan in panel (c). The source profile in Supplemental Fig. S14 is the same as that shown in Supplemental Fig. S13, providing a measure of the extent to which exciton hopping broadens the profile. The calculated profile shown in Supplemental Fig. S14 is consistent with the experimental result in terms of the approximate width of the profile. Taking $\mu \sim 1\,(\mathrm{ns})^{-1}$ gives $D \sim 1\,(\mathrm{cm})^2/\mathrm{s}$, consistent with the estimate of the exciton diffusion constant ($D \approx 0.5\,(\mathrm{cm})^2/\mathrm{s}$) made in the main text. In detail, however, the profiles do not match: small discrepancies can be seen in the tails – evident in the logarithmic plot of panel (b) – indicating that Eq. (S3) does not perfectly describe exciton motion on the dense substrate. In Supplemental Fig. S15 we show calculated profiles (green) for three different values of $D$ atop experimental data: the comparison indicates that the shape of the experimental profile is not perfectly described by exciton diffusion with a single diffusion constant.

The iso-energetic nature of our nanoparticles suggests that exciton subdiffusion of the type seen in previous work [3] and in Section S5 B does not occur in our experiments. Instead, the presence of imperfections in the nanoparticle substrate, and the practice of averaging experimental profiles over different spatial locations, may lead to non-diffusive profiles. To investigate this possibility we turned to microscopic simulations of the kind described in Section S5 A. On a square nanoparticle lattice we simulated the processes of creation (S19), self-destruction (S20), and hopping

$$\emptyset + A \overset{k}{\leftrightarrow} A + \emptyset, \qquad (S28)$$

using continuous-time Monte Carlo [4]. Hops were considered to any vacant nearest-neighbor nanoparticle. We assume the process of self-destruction to give rise to a photon at the same spatial location, and so the photon emission profile is the exciton profile $c(r)$ in the long-time limit. In Supplemental Fig. S16 we show example snapshots of the time-averaged photon emission statistics that result from these processes in the absence (left) and presence (right) of hopping. The source $\Phi(r)$ is again an Airy function of FWHM 240 nm; the left-hand panel is essentially an image of this function.

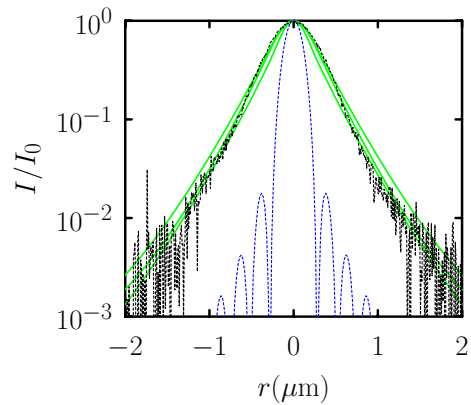

FIG. S15: As suggested by Supplemental Fig. S14, photoluminescence profiles resulting from the diffusion equation do not describe the shape of the experimental profile (black dashed line). The green lines show profiles $I(r) = (c \star S)(r)$, resulting from Eq. (S3) and Eq. (S18), for three different values of $D$.

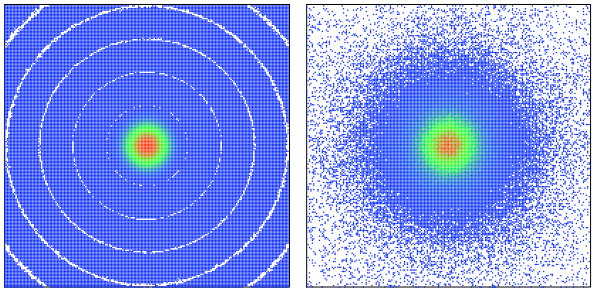

FIG. S16: Snapshots of time-averaged photon emission statistics from continuous-time Monte Carlo simulations of exciton creation and self-destruction on a square nanoparticle lattice. In the right-hand panel we also simulate exciton hopping. The laser source $\Phi(r)$ is an Airy function of FHWM 240 nm.

In Supplemental Fig. S17 we show the experimentally measured photoluminescence profile on a close-packed nanoparticle monolayer (black), together with profiles $I(r)$ (green) obtained by convolving, *via* Eq. (S18), the exciton profile $c(r)$ obtained from microscopic simulations with a point-spread Airy function $S(r)$ of FWHM 270 nm. We work in the low-power regime, with $\Phi_0/\mu = 10^{-2}$ (see Supplemental Fig. S9). Simulations done on pristine substrates match the diffusive profiles obtained using Eq. (S3). To mimic substrate imperfections we did microscopic simulations with a fraction $f_V$ of nanoparticle vacancies. No excitons can be created on, or hop to, a vacancy. We created vacancies in a spatially uncorrelated way, which is probably not true of vacancies produced by the nanoparticle self-assembly process: there, vacancies appear to cluster as gaps. However, the effect leads to profiles with shapes similar to those seen in experiment. The value of the hopping rate $k$ used to produce these simulations is $k/\mu = 10^3$. Taking $\mu \sim 1\,(\text{ns})^{-1}$ and the nanoparticle size $a \sim 10\,\text{nm}$ gives an estimate for the diffusion constant (on a pristine substrate) of $D \sim \frac{1}{2}10^3 \mu a^2 \sim \frac{1}{2}\,(\text{cm})^2 \text{s}^{-1}$. This value is consistent with the estimate derived from experimental data. The comparison of experimental and calculated profiles shown in Supplemental Fig. S17 suggests that energy transport in our experiments results from iso-energetic hopping, with a diffusion constant of order $(\text{cm})^2 \text{s}^{-1}$, on a spatially imperfect nanoparticle substrate. The middle curve in panel (a) of Supplemental Fig. S17, produced using a vacancy fraction of 20%, is the green line in Panel (g) of the main text.

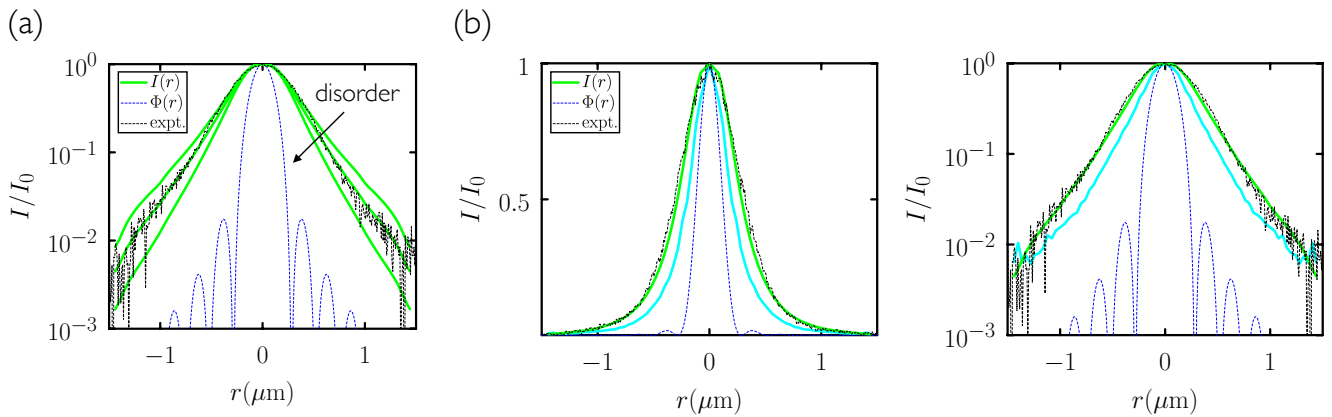

FIG. S17: Experimental photoluminescence profile (black dashed line) compared with simulated profiles $I(r) = (c \star S)(r)$ (green). The profile $c(r)$ results from continuous-time Monte Carlo simulations of exciton creation, self-destruction, and hopping; this function is convolved with $S(r)$, an Airy function of FWHM 270 nm, via Eq. (S18). In panel (a) we show results for parameters $\Phi_0/\mu = 10^{-2}$ and $k/\mu = 10^3$, with three different mean nanoparticle vacancy fractions $f_V$ of 0.1, 0.2, and 0.3 (from the outside in). In panels (b) and (c) we show results for vacancy fraction 0.2 on linear-linear and linear-log plots, respectively: its shape is a better match for the experimental profile than are profiles from the diffusion equation (see Figs. S14 and S15). The cyan lines are the exciton profiles $c(r)$.

## S7. ADDITIONAL EXPERIMENTAL FIGURES

- The excitation laser intensity profile is shown in Supplemental Fig. S18.

- The schematics of the optical setups are shown in Supplemental Fig. S19.

- Additional time-resolved PL spectroscopy data are provided in Supplemental Fig. S20.

- Additional lifetime measurements are provided in Supplemental Fig. S21.

- A time-resolved PL microscopy map is shown in Supplemental Fig. S22.

- The normalized PL intensity as a function of the excitation laser power is shown in Supplemental Fig. S23.

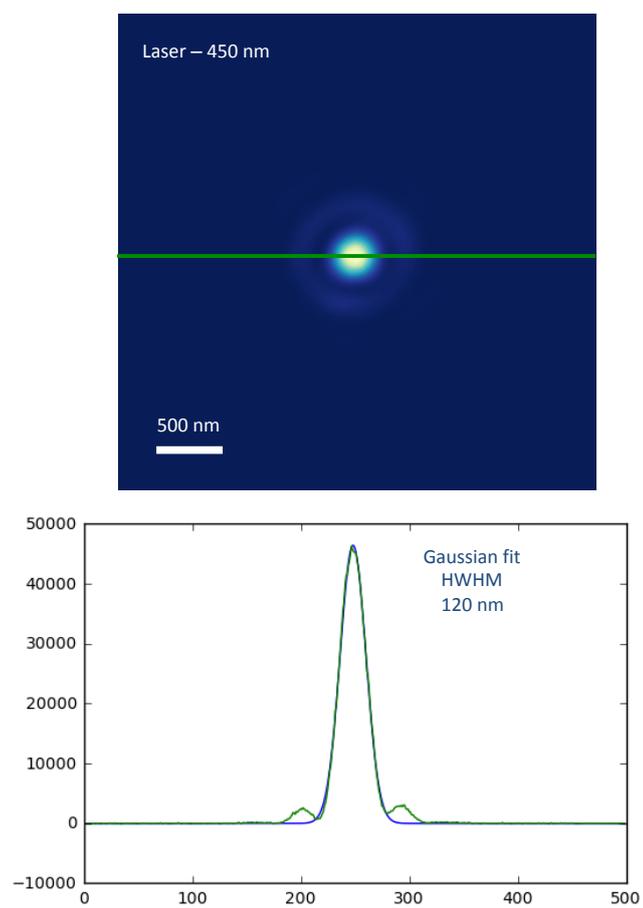

FIG. S18: (a) Diffraction limited CW laser spot (wavelength 450 nm) imaged with a CCD camera after 530X magnification. (b) Laser spot cross-section (blue) and Gaussian fit (FWHM 240 nm).

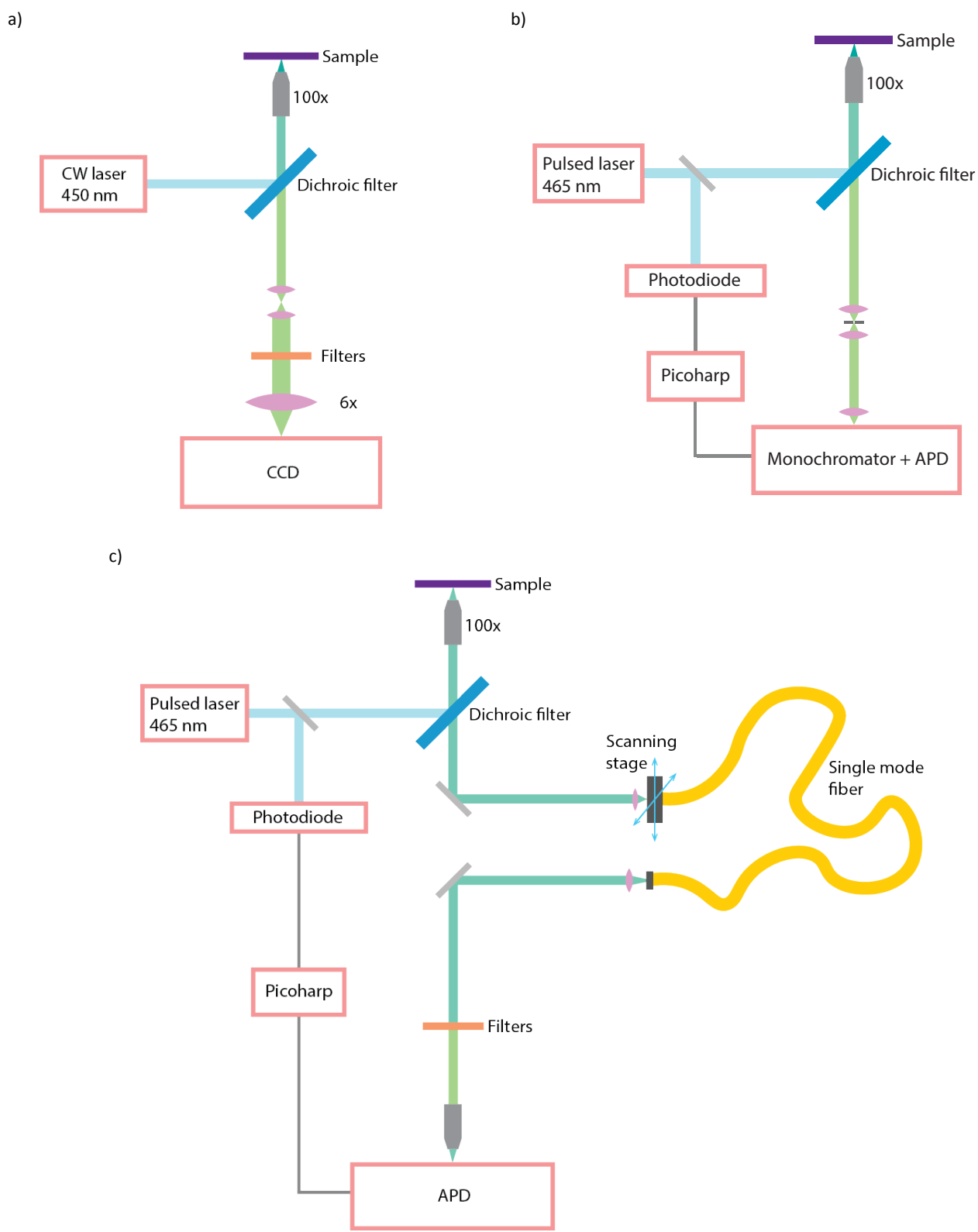

FIG. S19: Setup for (a) steady-state PL microscopy, (b) time-resolved PL spectroscopy, and (c) time-resolved PL microscopy.

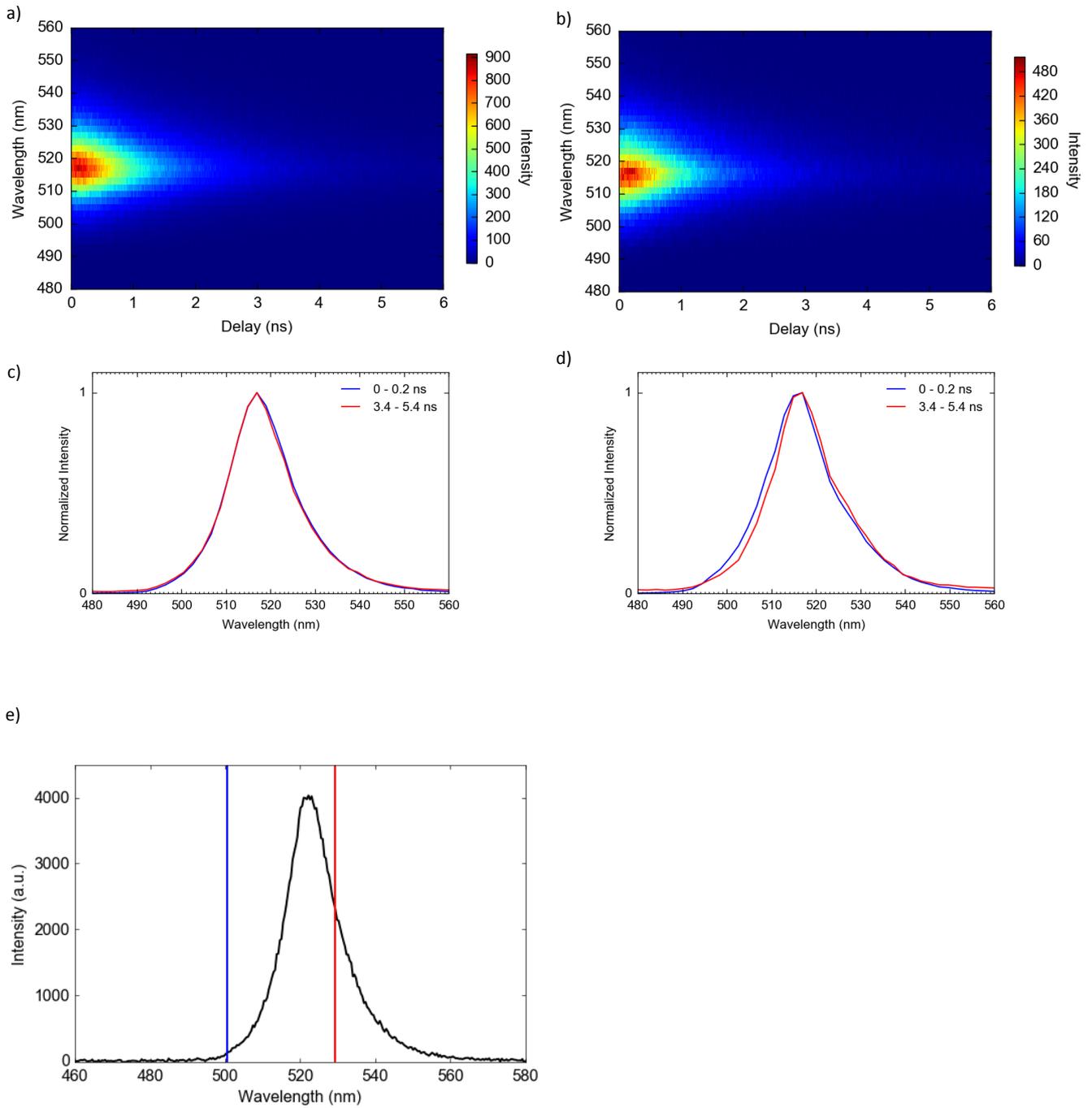

FIG. S20: Time-resolved PL spectra on (a) an ordered area made of uniformly sized PNCs and on (b) a disordered area made of PNCs of different sizes. (c) PL spectra from data in (a) integrated between 0 ns and 0.2 ns (blue) and between 3.4 ns and 5.4 ns (red). The two spectra overlap. (d) PL spectra from data in (b) integrated between 0 ns and 0.2 ns (blue) and between 3.4 ns and 5.4 ns (red). The spectrum at later time is slightly red shifted. (e) Integrated PL spectrum; the blue and red vertical lines show the PL wavelengths displayed in Figure 3-b and 3-c.

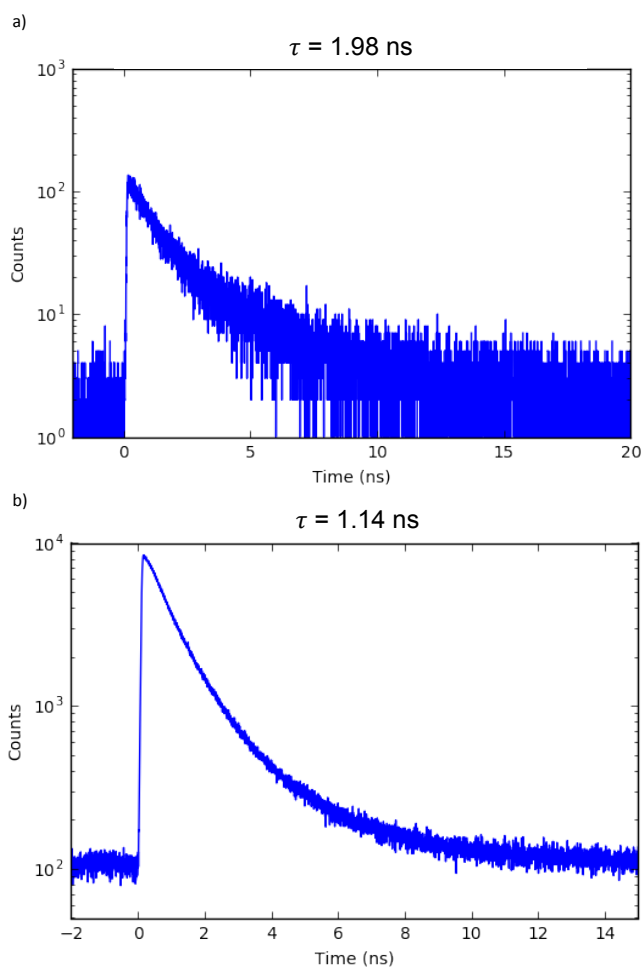

FIG. S21: Time-resolved PL of (a) PNCs assembled in a sparse monolayer; (b) PNCs assembled in a close-packed monolayer (integrated over the entire collection area). The lifetime of the system, measured as the time to reach a 37% or 1/e decay, is 1.94 ns, and 1.14 ns respectively.

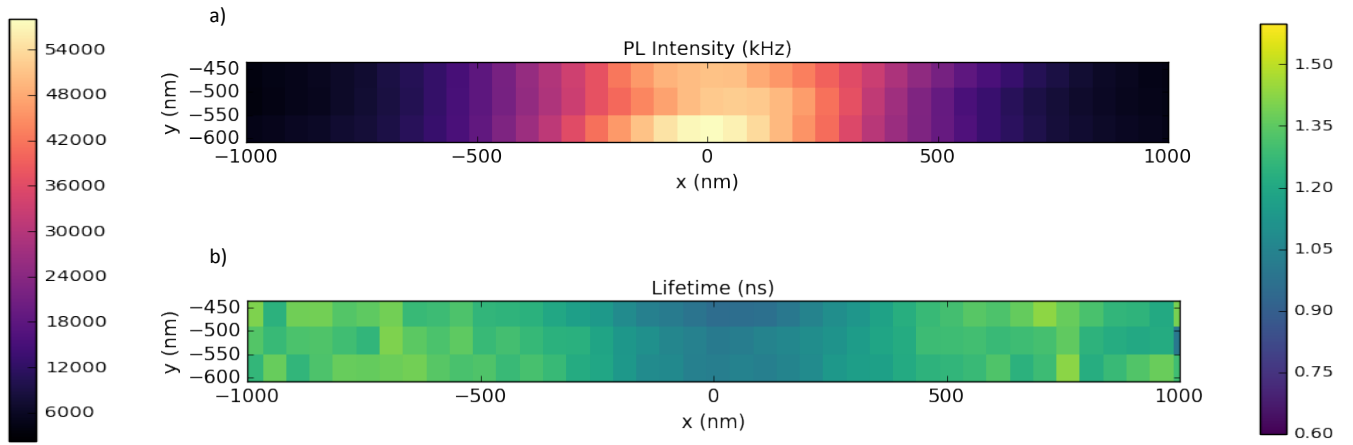

FIG. S22: (a) Time-integrated PL intensity map. (b) Lifetime map calculated from.

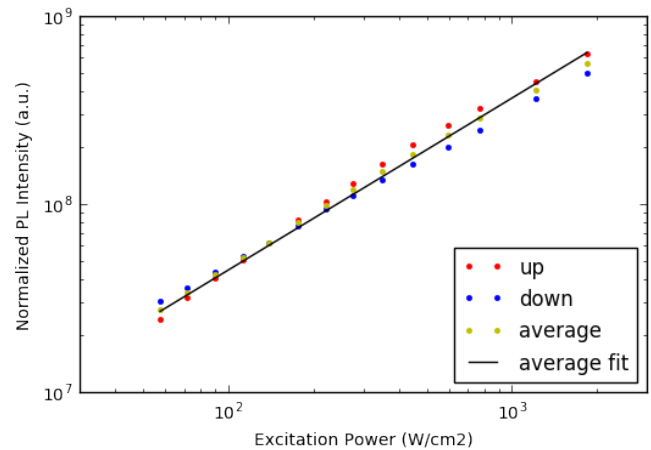

FIG. S23: Normalized PL intensity as a function of the excitation laser power. The upward scan is shown in red dots, the downward scan is shown in blue dots, and the average.